\documentclass[a4paper,fleqn]{cas-sc}

\usepackage[authoryear,longnamesfirst]{natbib}
\usepackage{xcolor}
\usepackage{amsmath,amssymb,amsfonts}
\usepackage{algorithmic}
\usepackage{graphicx}
\usepackage{textcomp}
\usepackage{subfigure}
\usepackage{arydshln}
\usepackage{epstopdf}
\usepackage{bm}
\usepackage{booktabs}

\usepackage{amsthm} 
\newtheorem{theo}{Theorem}
\newtheorem{lemm}{Lemma}
\newtheorem{defi}{Definition}
\newtheorem{rema}{Remark}
\newtheorem{assu}{Assumption}
\newtheorem{prob}{Problem}
\newtheorem{prop}{Proposition}

\def\tsc#1{\csdef{#1}{\textsc{\lowercase{#1}}\xspace}}
\tsc{WGM}
\tsc{QE}
\tsc{EP}
\tsc{PMS}
\tsc{BEC}
\tsc{DE}

\begin{document}
\let\WriteBookmarks\relax
\def\floatpagepagefraction{1}
\def\textpagefraction{.001}
\shorttitle{Bearing-based Simultaneous Localization and Affine Formation Tracking for Fixed-wing Unmanned Aerial Vehicles}
\shortauthors{H. Li et~al.}

\title [mode = title]{Bearing-based Simultaneous Localization and Affine Formation Tracking for Fixed-wing Unmanned Aerial Vehicles}                      



\author[1]{Huiming Li}[
                        orcid=0000-0002-5142-5151,
                        style=chinese]

\ead{huiminglhm@163.com}


\affiliation[1]{organization={College of Intelligence Science and Technology, National University of Defense Technology},
                city={Changsha},
                country={China}}

\author[2]{Zhiyong Sun}[style=chinese]
\ead{sun.zhiyong.cn@gmail.com}

\affiliation[2]{organization={Department of Mechanics and Engineering Science \& State Key Laboratory for Turbulence and Complex Systems, Peking University,~Beijing},
	country={China}
	}

\author[1,3]{Hao Chen}[%
   style=chinese
   ]
\cormark[1]
\ead{chenhao09@nudt.edu.cn}


\affiliation[3]{organization={Laboratory of Science and Technology on Integrated Logistics Support, National University of Defense Technology},
                city={Changsha},
                country={China}}

\author[1]{Xiangke Wang}[style = chinese]
\ead{xkwang@nudt.edu.cn}

\author[1]{Lincheng Shen}[style = chinese]
\ead{lcshen@nudt.edu.cn}

\cortext[cor1]{Corresponding author}


\nonumnote{This work was supported in part by the National Natural Science Foundation of China under Grant 62303483, U23B2032, U2241214, and 62376280, in part by the Research Project of the National University of Defense Technology under Grant ZK21-05.
  }

\begin{abstract}
This paper studies the bearing-based simultaneous localization and affine formation tracking (SLAFT) control problem for fixed-wing unmanned aerial vehicles (UAVs). In the considered problem, only a small set of UAVs, named leaders, can obtain their global positions, and the other UAVs only have access to bearing information relative to their neighbors.
To address the problem, we propose novel schemes by integrating the distributed bearing-based self-localization algorithm and the observer-based affine formation tracking controller. The designed localization algorithm estimates the global position by using inter-UAV bearing measurements, and the observer-based controller tracks the desired formation with the estimated positions. A key distinction of our approach is extending the SLAFT control scheme to the bearing-based coordination of nonholonomic UAV systems, where the desired inter-UAV bearings can be time-varying, instead of constant ones assumed in most of the existing results. Two control schemes with different convergence rates are designed to meet desired task requirements under different conditions. The stability analysis of the two schemes for SLAFT control is proved, and numerous simulations are carried out to validate the theoretical analysis.

\end{abstract}




\maketitle

\section{Introduction}
\label{sec:introduction}
Recently, robotics and control communities enthusiastically explore various control approaches for different multi-agent systems, such as unmanned aerial vehicles (UAVs) \cite{liu2019mission}, autonomous surface vessels (ASVs) \cite{Wen_TME_2024} and autonomous underwater vehicles (AUVs) \cite{Yao_AUV_2024}, to dramatically enhance efficiency and robustness.
Based on different types of available
measurements, most of the existing cooperative control laws can be divided into four categories: (i) position-based \cite{Zhang_position_2020,Wen_noncooperative_2022}; (ii) displacement-based~\cite{chen_formation_2021}; (iii) distance-based \cite{Mehdifar_distance_2020} and (iv) bearing-based \cite{zhao_bearingrigidity_2015}.
Among these measurements to achieve formation control, bearing measurements can be obtained by passive perception methods such as passive radars \cite{Gabard_2020} and cameras \cite{Li_3D_2022}, so that they have higher reliability, especially in complex electromagnetic environments or GPS-denied environments. 
With the wide application of vision sensors in robotics, formation control based on bearing measurements has attracted increasing attention due to the natural connection between the vision-based control problem and the bearing-based control strategies.

The primary foundation of bearing-based formation control is the so-called bearing rigidity theory \cite{zhao_bearingrigidity_2015}, which analyzes the conditions on multi-agent systems' topologies and physical configurations to ensure the uniqueness of the formation shape with inter-agent bearings.  
To further analyze the algebraic properties of bearing rigid formations, a matrix-weighted graph Laplacian, termed bearing Laplacian matrix, was proposed in \cite{zhao_localizability_2016}. With this powerful tool, a series of bearing-based, even bearing-only methods on formation control of single or double integrators were presented in  \cite{trinh_bearing-based_2018,li_bearing-only_2022,li_adaptive_2021,trinh_robust_2021}.
In \cite{van_tran_bearing-constrained_2022}, bearing-based control laws for unicycles were constructed to track moving leaders. 
However, to ensure the uniqueness, desired formations are always assumed to be infinitesimally bearing rigid, while only trivial infinitesimal bearing motions, namely translational and scaling motions, are the allowed formation maneuvers relying solely on the bearing rigidity.
Consequently, all the work mentioned above shares a common basic assumption: the desired inter-agent bearings are constant.

A few studies have focused on generating and tracking time-varying inter-agent bearings in multi-agent systems. Su et al. extended the properties of bearing rigidity to time-varying cases in \cite{Su_Bearing_2023}, where formation similarity maneuvers are achieved. Furthermore, Tang et al. investigated a new type of formation based on the persistence of excitation (PE) to relax the classical bearing rigidity conditions on the graph topology and achieve the generation and tracking for time-varying inter-agent bearings, which is called bearing persistently exciting (BPE) formation \cite{Tang_bearing-only_2020,Tang_Auto_2022,tang_localization_2023}. 
To guarantee the convergence of the proposed control protocols, the shape, scale or orientation of the desired formation is required to vary over time to avoid ambiguity in configurations. Moreover, the mentioned studies concentrate on linear systems, especially single- and double-integrator models. Due to the diversity of tasks and unmanned robots in practice, more exploration is needed to meet the various maneuvering needs of nonlinear systems (like fixed-wing UAVs) based on bearing measurements.

To extend the boundary of formation maneuvers, affine formation control problem was proposed and affine transformations were achieved in \cite{Lin_Affine_2016,2018_zhao_TAC}, which could deal with geometric motion transformations such as translation, rotation, scaling,  shearing and their combinations. 
Affine formation control strategies were applied with different dynamic models, including the linear time-invariant model \cite{Wu_affine_2023}, Euler-Lagrange model \cite{Li_LayeredAffine_2021}, and unicycle model \cite{LHM_FITEE}. The solution to dynamic affine formation tracking for six-degrees-of-freedom under-actuated multi-rotor  UAVs was considered in \cite{Xu_quad-rotor_2022}, where the desired three-dimensional target formation was maintained. Nevertheless, almost all research on affine formation control relies on the global position or displacement measurements, rather than easily available bearing information. The goal of our work is to integrate bearing measurements with affine transformations to improve the maneuverability of fixed-wing UAVs in perception-restricted environments, surpassing formation translation and scaling to achieve rotation and shearing in nonlinear formations.

To achieve cooperative tasks, simplified fixed-wing UAV models are established because the classic model described by 12-dimensional, state-coupled nonlinear equations \cite{beard2012small}, is too complex to design cooperative control law. 
A widely utilized model for fixed-wing UAVs is the two-dimensional nonholonomic unicycle model, where the flight altitude is assumed to be fixed. Various papers have proven its applicability in both simulations and flight experiments \cite{liu2019mission,HaoChen_path_2022,chen_formation_reconfiguration_2021}. Furthermore, by relaxing the assumption on fixed altitude, a three-dimensional nonlinear model can be used to describe the dynamic characteristics of fixed-wing UAVs as shown in \cite{Zhang_AST_2022}, which can depict UAVs' motion properties more realistically. The increase in complexity of 3D nonlinear models also adds the difficulty of designing control strategies. Therefore, there are few studies on the bearing-based affine formation control for three-dimensional nonholonomic UAV models.

Motivated by these observations, we investigate the bearing-based formation control problem of multiple fixed-wing UAVs with the simultaneous localization and affine formation tracking (SLAFT) control scheme. The time-varying inter-UAV bearings, determined by the relative position of UAVs, are utilized to estimate the positions of UAVs, which play a unique role in formation tracking control laws. Obviously, the difficulty of designing effective control schemes comes from two aspects. One is the nonlinearity of bearing measurements, and the formation configuration cannot be uniquely determined through bearings in certain special scenarios, e.g., two collinear formations with the same inter-UAV bearings but different configurations. The other is that the coupling system is not naturally compatible. That is, the stability of two modules, namely bearing-based localization and affine formation tracking, affects each other and more detailed theoretical analysis  needs to be conducted. Moreover, to describe the movement of fixed-wing UAVs, nonholonomic dynamics in three-dimensional space is considered in this paper, rather than the very simplified linear dynamics studied in \cite{ye_bearing-only_2017,Chen_2022_SLAF,tang_localization_2023}. 
The main contributions of this paper are summarized as follows:
\begin{itemize}
	\item[i)]  We establish the bearing-based SLAFT control scheme for fixed-wing UAVs, where the followers' positions are estimated by using the easily accessible bearing measurements and communication information obtained form their neighbors. Our proposed scheme allows nonholonomic UAVs to utilize portable sensors (cameras or directional sensor arrays) to obtain sufficient information to complete tasks, providing an alternative method in the perception-restricted environment (e.g., GPS-denied scenario). 
	\item[ii)] We propose two localization-and-tracking control schemes for affine formations to track not only constant formations but also time-varying ones. Comprehensive technical analysis is presented to validate the stability of the coupled system. Especially, a perturbation-based algorithm is introduced to deal with the dependence on the state-related persistent excitation condition while avoiding falling into unlocalizable configurations.
\end{itemize}

The remaining parts of this paper are organized as follows. We provide some mathematical preliminaries on graph theory and establish the bearing-based SLAFT control problem in Section~\ref{sec:ProblemState}. Theoretical results are proposed in Section~\ref{sec:main}. 
Section~\ref{sec:Simulation} provides simulations to validate the proposed schemes. Concluding remarks are given in Section~\ref{sec:Conclusion}, followed by appendices that contain the detailed proofs of the main results.

\section{PROBLEM DESCRIPTION} \label{sec:ProblemState}

\subsection{Preliminaries}
The interaction topology of a nonholonomic $n$-UAV system in this paper is represented by a fixed connected undirected graph $\mathcal{G} = \left(\mathcal{V}, \mathcal{E} \right)$, where $\mathcal{V} = \left\{1,2\cdots, n\right\}$ is the set of vertices, $\mathcal{E} \subset \mathcal{V} \times \mathcal{V}$ is the set of edges, and $|\mathcal{E}|=m$. 
An edge $\left(i,j\right) \in \mathcal{E}$ means there is a bidirectional interaction between UAV $i$ and $j$. 
Denote $n_l$ leaders as $\mathcal{V}_l =\lbrace 1, \cdots, n_l\rbrace $, and the 
remaining UAVs in the formation $n_f = n - n_l$ are regarded as followers denoted by $\mathcal{V}_f = \mathcal{V} \backslash \mathcal{V}_l$. 
The neighbor set of UAV $i$ is denoted by $\mathcal{N}_i := \left\{j \in \mathcal{V} \left|\left(i,j\right)\right. \in \mathcal{E}\right\}$. Hence, $\mathcal{N}_i = \mathcal{N}_i^l \cup \mathcal{N}_i^f $ where $\mathcal{N}_i^l := \left\{j \in \mathcal{V}_l \left|\left(i,j\right)\right. \in \mathcal{E}\right\}$ and $\mathcal{N}_i^f := \left\{j \in \mathcal{V}_f \left|\left(i,j\right)\right. \in \mathcal{E}\right\}$.
Let $\bm{p}_i\in\mathbb{R}^d$ be the position of UAV $i$, and $\bm{p}:=\left[\bm{p}_l^T,\bm{p}_f^T\right]^T=\left[\bm{p}_1^T,\cdots,\bm{p}_n^T\right]^T \in \mathbb{R}^{dn}$, where the superscript $d=2, 3$ denotes the spatial dimension.
A \emph{formation} can be expressed by $\left(\mathcal{G},\bm{p}\right)$ if the vertex $i$ in $\mathcal{G}$ is mapped to $\bm{p}_i$ for all $i \in \mathcal{V}$.

An oriented graph is constructed when assigning a direction to each edge in the undirected graph $\mathcal{G}$. 
The incidence matrix $\bm{H} \in \mathbb{R}^{m\times n}$ is a matrix whose elements is composed of $1, ~0$ and $-1$, describing the edge direction of the oriented graph. The rows of an incidence matrix are indexed by edges and columns are labeled by vertices. $\left[\bm{H}\right]_{ki} = 1$ if vertex $i$ is the head of edge $k$,  $\left[\bm{H}\right]_{ki} = -1$ if $i$ is the tail, and $\left[\bm{H}\right]_{ki} = 0$ otherwise. 

Assign a set of scalars to the edges of $\mathcal{G}$, e.g., $\varpi_{ij} \in \mathbb{R}$ for $ (i,~j)\in \mathcal{E}$ and $\varpi_{ij}=\varpi_{ji} $, then these scalars can be seen as the \emph{stress}. An \emph{equilibrium stress} is established if and only if $\sum_{j \in \mathcal{N}_{i}} \varpi_{i j}\left(\bm{p}_{j}-\bm{p}_{i}\right)=\bm{0}$ holds in the formation $(\mathcal{G},~\bm{p})$, which can be rearranged as $\left(\bm{\Omega} \otimes \bm{I}_d\right)\bm{p} =\bm{0}$, where the symbol $\otimes$ represents the Kronecker product and the \emph{stress matrix} $\bm{\Omega} \in \mathbb{R}^{n\times n}$ is defined as below.
\begin{align}
	[\bm{\Omega}]_{i j}=\left\{\begin{array}{ll}
		-\varpi_{i j}, & \text { if } i \neq j \text { and } j \in \mathcal{N}_{i} \\
		0, & \text { if } i \neq j \text { and } j \notin \mathcal{N}_{i} \\
		\sum_{k \in \mathcal{N}_{i}} \varpi_{i k}. & \text { if } i=j
	\end{array}\right.\notag
\end{align}
The matrix $\boldsymbol{I}_d$ is the identity matrix. 
Let $\bar{\bm{\Omega}}:=\bm{\Omega} \otimes \bm{I}_d=\left[\begin{array}{cc}
	\bar{\bm{\Omega}}_{l l} & \bar{\bm{\Omega}}_{l f} \\
	\bar{\bm{\Omega}}_{f l} & \bar{\bm{\Omega}}_{f f}
\end{array}\right]$, 
where $\bar{\bm{\Omega}}_{ll} \in\mathbb{R}^{(dn_l)\times (dn_l)}$, $\bar{\bm{\Omega}}_{lf} \in\mathbb{R}^{(dn_l)\times (dn_f)},~\bar{\bm{\Omega}}_{fl} \in\mathbb{R}^{(dn_f)\times (dn_l)}$, and $\bar{\bm{\Omega}}_{ff} \in\mathbb{R}^{(dn_f)\times (dn_f)}$.

\subsection{Affine Formation Control}
Due to the variety of affine transformations, the affine formation control scheme has a great advantage in enhancing formation maneuverability. We define a constant \emph{nominal configuration} $\bm{r}$ in $\mathbb{R}^{d}$, where $\bm{r}=\left[\bm{r}_{\ell}^{T}, \bm{r}_{f}^{T}\right]^{T} =\left[\bm{r}_{1}^{T}, \ldots,\bm{r}_{n}^{T}\right]^{T}$. The \emph{nominal formation} is denoted by  $\left(\mathcal{G},~\bm{r}\right)$, and the time-varying target formation is affinely transformed from $\bm{r}$ as follows.
\begin{defi}[Target Formation] \label{definition:target_formation}
	The time-varying desired position for all UAVs is expressed as
	\begin{equation}
		\bm{p}^{*}(t)=\left[\bm{I}_d \otimes \bm{A}(t)\right] \bm{r}+\bm{1}_d \otimes\bm{ b}(t) ,
	\end{equation}
	where $\bm{p}^*(t)=\left[\bm{p}_1^{*T}(t),\cdots,\bm{p}_n^{*T}(t)\right]^T$ and $\bm{1}_d= \left[1,\cdots,1\right]^T$. The invertible matrix $\bm{A}(t)\in \mathbb{R}^{d\times d}$ and translation vector $\bm{b}(t)\in \mathbb{R}^d$ are continuous on $t$. Accordingly, the $i$-th UAV's desired position in the target formation $\left(\mathcal{G},\bm{p}^*\right)$ is denoted by $\bm{p}_i^{*}(t)=\bm{A}(t) \bm{r}_i+ \bm{b}(t)$.
\end{defi}

An important requirement of the affine formation control method is \emph{affine localizability}, that is, the desired formation can be determined by leaders in the nominal formation $\left(\mathcal{G},\bm{r}\right)$. 
\begin{defi}[Affine Localizability \cite{2018_zhao_TAC}]\label{definition:local}
	The nominal formation $(\mathcal{G},~\bm{r})$ is affinely localizable by the leaders if for any $\bm{p}=\left[\bm{p}_l^T,\bm{p}_f^T\right]^T \in \mathcal{A}\left(\bm{r}\right)$, $\bm{p}_f$ can be uniquely determined by $\bm{p}_l$, where $$\mathcal{A}\left(\bm{r}\right) := \left\{\bm{p}\in \mathbb{R}^{dn}: \bm{p}=\left[\bm{I}_d \otimes \bm{A}\right] \bm{r}+\bm{1}_d \otimes\bm{ b}, \forall \bm{A}\in  \mathbb{R}^{d\times d}, \bm{b}\in \mathbb{R}^d\right\}$$ contains all the affine transformations of $\bm{p}$.
\end{defi}

To select proper leaders and ensure affine localizability, we  introduce a term \emph{affine span} $\mathcal{S}$ as $$\mathcal{S}:=\left\{\sum_{i=1}^{n} a_i \bm{p}_i : a_i \in \mathbb{R}~\text{for all}~i,~\text{and}~ \sum_{i=1}^{n} a_i=1 \right\}$$
It is clarified that at least $d+1$ leaders are needed to affinely span $\mathbb{R}^d$, which requires $n_l \ge d+1$ \cite{2018_zhao_TAC}. 
Subsequently, the following assumption is imposed to construct an affine formation control scheme, and the specific theoretical analysis can be found in \cite{2018_zhao_TAC,LHM_FITEE}.

\begin{assu}\label{assump:nominal}
	The following conditions exist with the nominal formation $(\mathcal{G},~\bm{r})$:
	\begin{itemize}
		\item[(i)] For $(\mathcal{G},~\bm{r})$, $\left\{\bm{r}_i\right\}_{i=1}^n$ affinely span $\mathbb{R}^d$; 
		\item[(ii)] The stress matrix $\bm{\Omega}$ of $(\mathcal{G},~\bm{r})$ is positive semi-definite and $\operatorname{rank}(\bm{\Omega})=n-d-1$;
		\item[(iii)] The nominal formation $(\mathcal{G},~\bm{r})$ is affinely localized by the selected leaders.
	\end{itemize}
\end{assu}

With Assumption~\ref{assump:nominal}, it can be proved that $\bar{\bm{\Omega}}_{ff}$ is positive definite. Meanwhile, the desired position of followers $\bm{p}^*_f$ can be uniquely calculated by $\bm{p}^*_l$, i.e., $\bm{p}_f^*=-\bar{\bm{\Omega}}_{ff}^{-1}\bar{\bm{\Omega}}_{fl}\bm{p}_l^*$  \cite{2018_zhao_TAC}.

\subsection{Problem Formulation}
In this paper, we study the bearing-based SLAFT problem for fixed-wing UAVs to achieve affine transformations, since bearing measurement is easily acquired even if global position is not available in the perception-restricted environment. In the considered scenario, there are $n_l$ leaders with known initial positions at a global coordinate frame (also can be seen as \emph{anchors}) equipped with high-precision inertial measurement units to obtain their global positions, and the remaining fixed-wing UAVs can only measure the bearing vectors with respect to their neighbors by using cameras. Denote the edge and bearing vectors between UAV $i$ and $j$ as $\bm{e}_{ij} := \bm{p}_j-\bm{p}_i$ and $\bm{g}_{ij}:=\frac{\bm{e}_{ij}}{\|\bm{e}_{ij}\|}$.
If edge $k$ in the underlying graph $\mathcal{G}$ connects UAV $i$ and $j$, we can write $\bm{e}_k :=\bm{e}_{ij}$. The corresponding bearing vector is then denoted by $\bm{g}_k := \frac{\bm{e}_k}{\|\bm{e}_k\|}$. Let $\bm{\pi}_{g_k} := \bm{I}_d -\bm{ g}_k \bm{g}_k^T$ and $\bm{\Pi} := \operatorname{diag}(\bm{\pi}_{g_k})\in \mathbb{R}^{dm \times dm}$, then the bearing Laplacian matrix is introduced as follows:
\begin{equation}
	\bm{L}_B = \bm{\bar{H}}^T\bm{\Pi}\bm{\bar{H}} \in \mathbb{R}^{dn \times dn},
\end{equation}
where $\bm{\bar{H}}:=\bm{H }\otimes \bm{I}_d $.
It is easy to obtain that the bearing Laplacian matrix $\bm{L}_B $ is symmetric positive semi-definite and $\operatorname{span}\left(\bm{1} \otimes \bm{I}_d, \bm{p}\right) \subseteq \operatorname{Null}\left(\bm{L}_B\right)$  \cite{zhao_localizability_2016}. Rewrite $\bm{L}_B$ into the block matrix form as $\bm{L}_B =\left[\begin{array}{cc}
	\bm{L}_{Bll} & \bm{L}_{Blf}\\
	\bm{L}_{Bfl} & \bm{L}_{Bff}
\end{array}\right]$ where $\bm{L}_{Bfl} \in \mathbb{R}^{dn_f \times dn_l}$ and $\bm{L}_{Bff} \in \mathbb{R}^{dn_f \times dn_f}$.

\begin{lemm}[Bearing Localizability \cite{zhao_localizability_2016}] \label{lemma:localizable}
	For a static multi-agent network, one has (i) $\bm{L}_{Bff}\bm{p}_f = -\bm{L}_{Bfl}\bm{p}_l$; (ii) The network is localizable ($\bm{p}_f $ can be uniquely determined) if and only if $\bm{L}_{Bff}$ is nonsingular; (iii) If the network is localizable, then $\bm{p}_f$ can be uniquely calculated by $\bm{p}_f = -\bm{L}_{Bff}^{-1} \bm{L}_{Bfl}\bm{p}_l$.
\end{lemm}
\begin{rema}
	Lemma~\ref{lemma:localizable} reveals the relationship between the localizability and the nonsingularity of bearing Laplacian submatrix $\bm{L}_{Bff}$ involving all followers. Anchors (or leaders) are necessary for the bearing-based localization algorithms to eliminate the ambiguity caused by the lack of position or distance information. At least one distance is required to remove the scale uncertainty \cite{su_bearing-based_2023}. 
	\cite{zhao_localizability_2016} argue that the number of anchors $n_a$ should satisfy $n_a\ge \frac{\operatorname{dim}\left(\operatorname{Null}\left(\bm{L}_B\right)\right)}{d} >1$. 
\end{rema}
\begin{rema}
	Compared with classical position measurements, bearings have potential application prospects in some perception-restricted environments. Taking the GPS-denied environment as an example, bearing-based methods provide alternative solutions for the multi-agent system to maintain its ability to perform various tasks while reducing the perception cost.
\end{rema}

It is assumed that the sensing topology is described by a connected undirected graph, which implies that each UAV $i \in \mathcal{V}$ can communicate with its neighbor $j \in \mathcal{N}_i$ to exchange information, such as bearing measurements. Moreover, UAVs are assumed to get their orientations relative to the inertial coordinate system through sensors such as gyroscopes.

In the considered scenario, fixed-wing UAVs are expected to form a desired formation $\left(\mathcal{G},\bm{p}^*(t)\right)$, as described in Definition~\ref{definition:target_formation}. We impose the following assumptions regarding the desired formation.
\begin{assu} \label{assu:Dp}
	The desired position $\bm{p}_i^*(t)$ is chosen to meet the following conditions:
	\begin{itemize}
		\item[(1)] $\bm{p}_i^*(t)$ is twice differentiable with respect to $t$;
		\item[(2)] $\bm{p}_i^*(t)$ and $\dot{\bm{p}}_i^*(t)$ are bounded for all $t$, and the desired relative distance $\|\bm{e}_{ij}^*(t)\|_{j \in \mathcal{N}_i} >0$ so that the desired bearing $\bm{g}_{ij}^*(t)$ is well-defined.
	\end{itemize}
\end{assu}
In this work, we focus on the formation tracking control problem, where we assume leaders are fully controlled for clarity, i.e., $\bm{p}_i^*(t) = \bm{p}_i(t)$ for $i \in \mathcal{V}_l$. The control of leaders is beyond the scope of this paper, but it can be achieved by various distributed consensus protocols, like position-based algorithms \cite{Li_LayeredAffine_2021,Fang_TAC_2024,chen_formation_2021}. Moreover, the leaders' positions are assumed to be known by themselves.
To achieve the control objectives, two problems are raised that need to be solved, as shown below.
\begin{prob}[Bearing-based Localization]\label{prob1:localization}
	Consider a team of fixed-wing UAVs, while only leaders have access to global positions. Design a localization algorithm to estimate the global position for followers by using inter-neighbor bearing information $\bm{g}_{ij}(t)$, such that, $\hat{\bm{p}}_i(t) \to \bm{p}_i(t)$ for $i \in \mathcal{V}_f$, where $\hat{\bm{p}}_i(t)$ is the estimated position of the $i$-th UAV. 
\end{prob}
\begin{prob}[Affine Formation Tracking]\label{prob2:FormationTracking}
	Under Assumptions~\ref{assump:nominal}-\ref{assu:Dp}, design control laws for followers to track a time-varying target formation $\left(\mathcal{G},\bm{p}^*(t)\right)$, and achieve the desired affine transformations with the bearing-based localization algorithm, such that $\bm{p}_i(t) \to \bm{p}_i^*(t)$ and $\bm{g}_{ij}(t) \to \bm{g}_{ij}^*(t)$ for the $i$-th ($i \in \mathcal{V}_f$) fixed-wing UAV when $t \to \infty$.
\end{prob}

It is worth noting that Problems~\ref{prob1:localization} and \ref{prob2:FormationTracking} are not decoupled, but the localization and affine formation tracking should be conducted simultaneously. 

	\section{Main Results}\label{sec:main}

\begin{figure}
	\centering
	\includegraphics[scale=0.3]{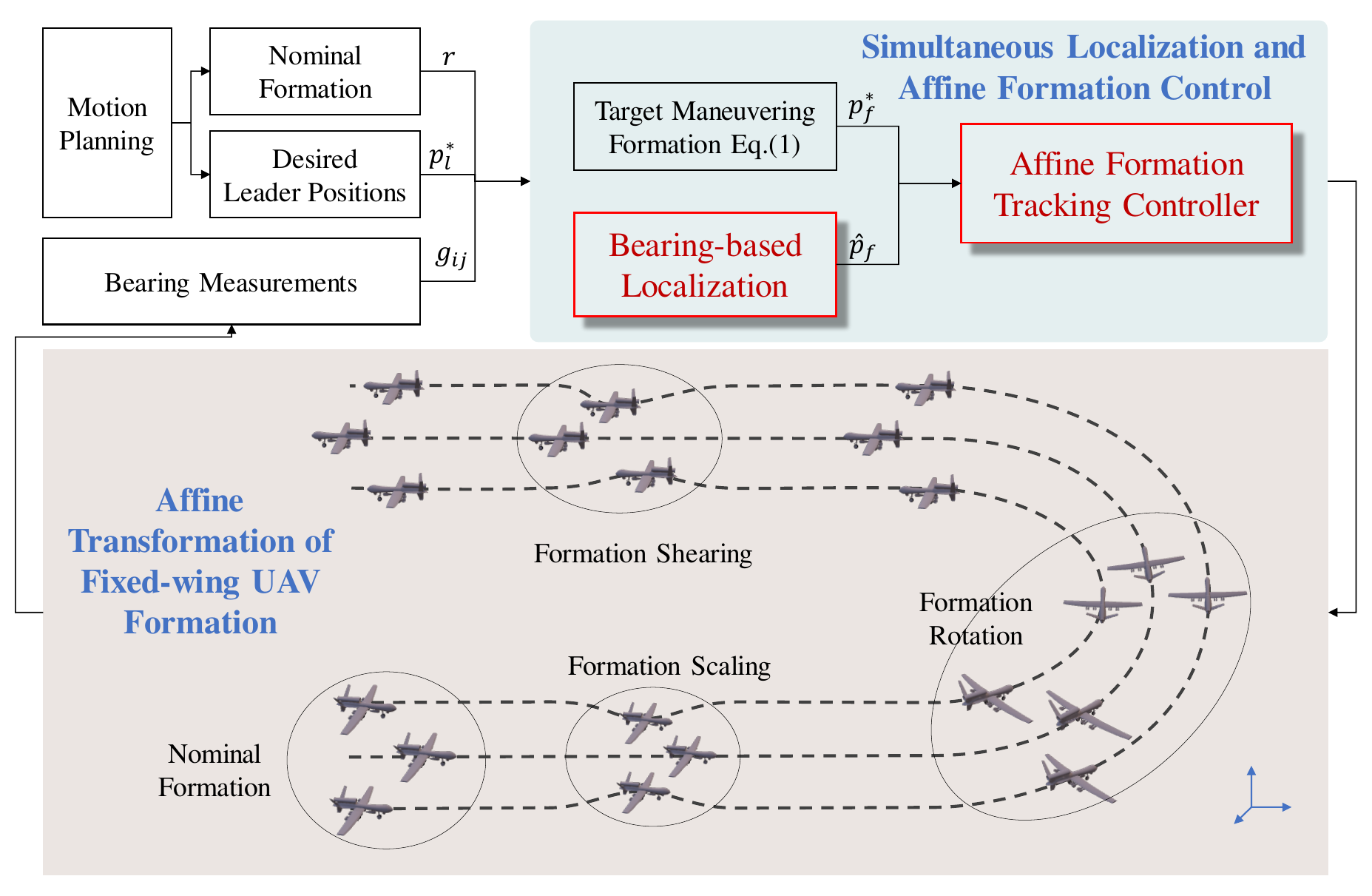} 
	\caption{An illustration of bearing-based simultaneous localizationa and affine formation control scheme.}
	\label{fig:illustration}
\end{figure}

To handle Problems~\ref{prob1:localization} and \ref{prob2:FormationTracking}, the SLAFT control scheme is designed in this section, as illustrated in  Fig.~\ref{fig:illustration}. Trajectories are generated with motion planning techniques, and the nominal formation is preordained. Bearing measurements are utilized to estimate each UAV's position, and the estimated position is applied to realize the affine formation tracking. Consequently, target affine transformations, including rotation, shearing, scaling and their combinations, resulting in constant or time-varying inter-UAV bearings, are realized with the proposed bearing-based SLAFT control protocols. Significantly, the localization process and affine formation tracking process are not naturally compatible, and their performance depends on each other, which increases the complexity of analyzing the characteristics of the coupled nonlinear system.

\subsection{The Fixed-wing UAV Model}

The dynamic model of the $i$-th fixed-wing UAV in three-dimensional space is given as follows \cite{beard2012small,Zhang_AST_2022,Yang_2021_3D}, 
\begin{equation}\label{eq:3d_unicycle}
	\left\{\begin{array}{l}
		\dot{x}_i = v_i \cos\alpha_i \cos\beta_i,\\
		\dot{y}_i = v_i \cos\alpha_i \sin\beta_i,\\
		\dot{z}_i = v_i \sin\alpha_i, \\
		\dot{v}_i = -g \sin\alpha_i + \dfrac{T_i-F_i}{m_i},\\
		\dot{\alpha}_i =  \dfrac{1}{m_i v_i}\left(L_i\cos\varrho_i-m_i  g\cos\alpha_i\right),\\
		\dot{\beta}_i = \dfrac{L_i \sin \varrho_i}{m_iv_i\cos \alpha_i},
	\end{array}\right. 
\end{equation}
where $\bm{p}_i = \left[x_i,y_i,z_i\right]^T$ is the position of the $i$-th fixed-wing UAV expressed in the inertial coordinate system; $v_i$ is the speed of the UAV; $m_i$ and $g$ are the mass and gravity acceleration, respectively; $\alpha_i$ is the flight path angle, $\beta_i$ is the heading angle, and $F_i$ is the air friction. The control inputs of the $i$-th UAV are total thrust $T_i$, the lift force $L_i$ and the banking angle $\varrho_i$. Generally, the engine thrust $T_i$ can be controlled by throttle, the lift force $L_i$ is generated by the elevator, and the banking angle $\varrho_i$ is controlled by the rudder and aileron.

\begin{figure}
	\centering
	\includegraphics[scale=0.25]{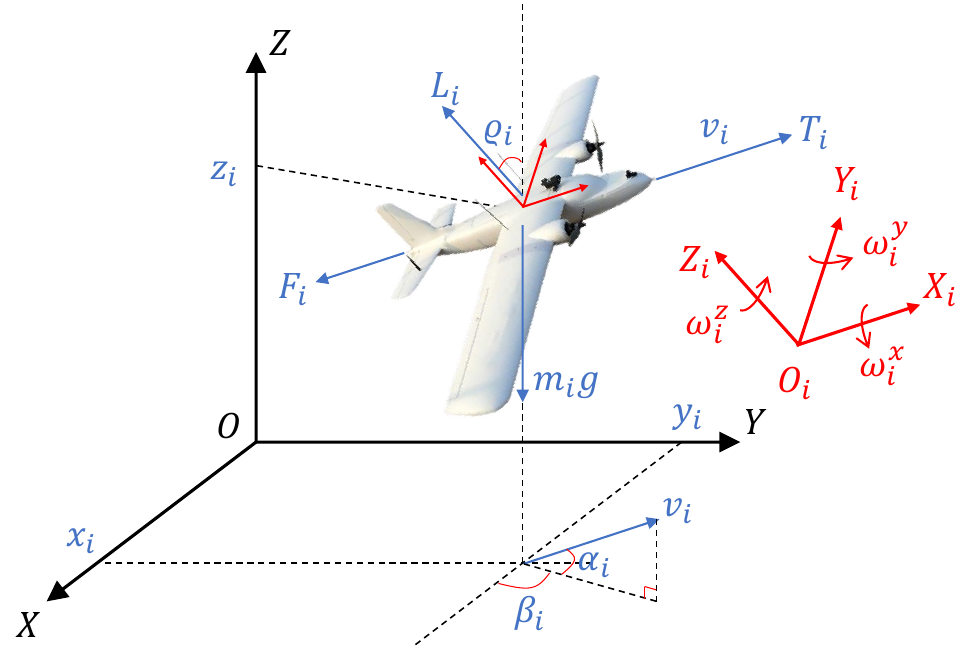} 
	\label{fig:3D_UAV}
	\caption{Model Description of Fixed-wing UAVs in three-dimensional space.}
\end{figure}

\begin{rema}
	Eq.~\eqref{eq:3d_unicycle} is derived from 12-dimensional, state-coupled nonlinear dynamics with certain simplifying assumptions. The main assumption is that the angles of attack and sideslip are negligible. More technical details can be found in Section 9 of \cite{beard2012small}. The popularity of Eq.~\eqref{eq:3d_unicycle} is due to the fact that it reduces the mathematical complexity  in designing guidance control laws, while still capturing the essential behavior of UAV dynamics.
	
\end{rema}

Let $\chi_i = -\alpha_i$. The attitude of the $i$-th UAV can be described by $\bm{R}_i $, where
\begin{equation}
	\bm{R}_i=\left[\begin{array}{ccc}
		\cos\beta_i \cos\chi_i & - \sin \beta_i  & \cos \beta_i \sin \chi_i\\
		\sin \beta_i \cos \chi_i & \cos \beta_i & \sin \beta_i \sin \chi_i\\
		-\sin \chi_i & 0 & \cos \chi_i
	\end{array}\right]
\end{equation}
and $\bm{R}_i^T = \bm{R}_i^{-1}$. The kinematics of the $i$-th UAV's attitude is 
\begin{equation}\label{eq:Dot_Rotation}
	\dot{\bm{R}}_i = \bm{\omega}_i^\wedge  \bm{R}_i ,
\end{equation}
where $(\cdot)^{\wedge}$ defines a map from a vector to a skew symmetric matrix in $\mathbb{R}^{3 \times 3}$, $\bm{\omega}_i = \left[\omega_i^x,\omega_i^y,\omega_i^z\right]^T$ is the angular velocity of the $i$-th UAV expressed in the inertial frame, and $\omega_i^x = -\dot{\chi}_i \sin \beta_i$, $\omega_i^y =\dot{\chi}_i \cos \beta_i$, $\omega_i^z = \dot{\beta}_i $. Thus,
\begin{equation}
	\bm{\omega}_i^\wedge = \left[\begin{array}{ccc}
		0 & -\dot{\beta}_i  &\dot{\chi}_i \cos \beta_i\\
		\dot{\beta}_i & 0 & \dot{\chi}_i \sin \beta_i\\
		-\dot{\chi}_i \cos \beta_i & -\dot{\chi}_i \sin \beta_i & 0
	\end{array}\right].
\end{equation}

\begin{rema}
	Given the matrix $\bm{R}_i$, the matrices $\dot{\bm{R}}_i \bm{R}_i^{-1} \in \mathbb{R}^{3 \times 3} $ and $\bm{R}_i^{-1}  \dot{\bm{R}}_i \in \mathbb{R}^{3 \times 3} $ are both skew symmetric. Thus, the velocity of a rotating body can be represented by $\left(\bm{\omega}_i\right)^\wedge:= \dot{\bm{R}}_i \bm{R}_i^{-1}$ or $\left(\bm{\omega}_i^b\right)^\wedge:=\bm{R}_i^{-1}  \dot{\bm{R}}_i $, where $\bm{\omega}_i$ is defined in the inertial frame and $\bm{\omega}_i^b$ is expressed in the body frame \cite{Murray_Robotic_1994}. In our paper, the inertial coordinate system is assumed to be available. Accordingly, Eq.~\eqref{eq:Dot_Rotation} is established for the rotational dynamics, inspired by \cite{Van_TCNS_2020,van_tran_bearing-constrained_2022}. Denote the first column of $\bm{R}_i$ as $\bm{h}_i$, the heading vector of UAV $i$, then we have $\dot{\bm{h}}_i =\bm{\omega}_i^\wedge \bm{h}_i $ and 
	$-\left(\bm{h}_i^\wedge\right)^2 = \bm{I}_3 - \bm{h}_i \bm{h}_i^T$.
\end{rema}

Therefore, we can reformulate the model Eq.~\eqref{eq:3d_unicycle} as the following equivalent model, 
\begin{equation}\label{eq:3D_unicycle_matrix}
	\left\{\begin{array}{l}
		\dot{\bm{p}}_i = \bm{R}_i \bm{u}_i,\\
		\dot{\bm{h}}_i = \bm{\omega}_i^\wedge \bm{h}_i, \\
		\dot{v}_i = \tau_i,
	\end{array}\right. 
\end{equation}
where $\bm{u}_i = \left[v_i,0,0\right]^T$. The variable $\tau_i$ represents the virtual thrust, a control input of system \eqref{eq:3D_unicycle_matrix}. Another control input is the angular speed $\bm{\omega}_i$. The real inputs in model \eqref{eq:3d_unicycle} can be deduced from $\tau_i$ and $\bm{\omega}_i$ as follows.
\begin{align}
	T_i =& m_i \tau_i + F_i - m_ig \sin \chi_i , \label{eq:3D_Trans1}
	\\
	L_i =& m_i \sqrt{\left( v_i  \cos \chi_i \omega_i^z\right)^2+  \left( g \cos \chi_i - v_i \dot{\chi}_i \right)^2} , \label{eq:3D_Trans2}
	\\
	\varrho_i = & \tan^{-1}\left(\dfrac{v_i \cos \chi_i \omega_i^z}{g \cos \chi_i-v_i \dot{\chi}_i }\right). \label{eq:3D_Trans3}
\end{align}

Then, we provide detailed presentations and analysis to our proposed SLAFT control strategies.

\subsection{The SLAFT control scheme}

We aim to simultaneously solve Problems \ref{prob1:localization}-\ref{prob2:FormationTracking} for the followers. We suppose that leaders move with constant velocities, i.e., $\tau_i = 0$ for $i \in \mathcal{V}_l$, while $\bm{\omega}_r $ and $ \bm{h}_r$ are constant angular velocity and heading vector for leaders.	
By assuming that UAVs can communicate their position estimations with their neighbors, 
we design the SLAFT algorithm for the followers as follows:
\begin{subequations}\label{eq:SLAF_ori}
	\begin{align}
		\dot{\hat{\bm{p}}}_i =& -k_\delta \sum_{j \in \mathcal{N}_i^f}\bm{\pi}_{\bm{g}_{ij}}\left(\hat{\bm{p}}_i-\hat{\bm{p}}_j\right) -k_\delta \sum_{j \in \mathcal{N}_i^l}\bm{\pi}_{\bm{g}_{ij}}\left(\hat{\bm{p}}_i-\bm{p}_j\right)  + \dot{\bm{p}}_i \label{eq:pos_esti} ,\\
		\tau_i =& -\bm{h}_i^T \bm{f}_i\label{eq:3D_AffineCtrl1_1},\\
		\bm{\omega}_i =& - \dfrac{1}{v_i} \bm{h}_i \times \bm{f}_i + \bm{\aleph}_i \label{eq:3D_AffineCtrl1_2},\\
		\bm{f}_i =&  \sum_{j \in \mathcal{N}_i^f}k_p \varpi_{ij}  \left(\hat{\bm{p}}_i-\hat{\bm{p}}_j\right)+  \sum_{j \in \mathcal{N}_i^l}k_p \varpi_{ij}  \left(\hat{\bm{p}}_i-\bm{p}_j\right)+ \sum_{j \in \mathcal{N}_i} k_v \varpi_{ij}  \left(\bm{h}_iv_i - \bm{h}_jv_j \right),
	\end{align}
\end{subequations}
where $k_\delta$, $k_p$ and $k_v$ are positive coefficients. The vector $\bm{\aleph}_i$ satisfies $\bm{\aleph}_i^\wedge \bm{h}_i = \dfrac{v_i^*}{v_i} \bm{\omega}_r^\wedge \bm{h}_r $ for $i =\left\{n_l+1, \cdots, n\right\}$, which serves as a feedforward term for the UAV's acceleration, assisting to eliminate formation tracking errors.
The variable $\dot{\bm{p}}_i$ represents the velocity of the $i$-th UAV, which can be measured by airborne sensors, independent of global position information. 

\begin{rema}
	\begin{figure}
		\centering
		\includegraphics[scale=0.3]{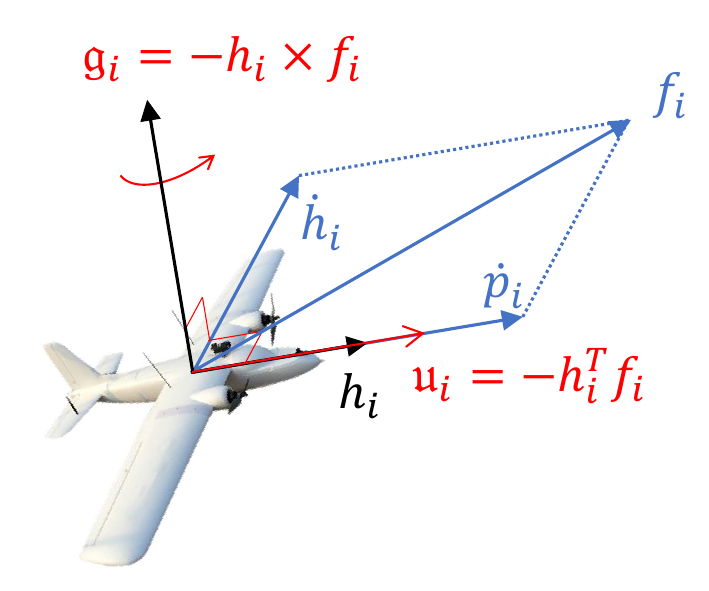} 
		\caption{An illustration of projection-inspired control scheme for fixed-wing UAVs in three-dimensional space.}
		\label{fig:3D_Design}
	\end{figure}
	
	Inspired by \cite{Zhao_General_2018}, the significance of formation tracking control laws \eqref{eq:3D_AffineCtrl1_1}-\eqref{eq:3D_AffineCtrl1_2} is illustrated in Fig.~\ref{fig:3D_Design}. The control item $\bm{f}_i$ is designed to describe the disparities between current states and the desired ones, 
	while leading the evolution of the whole multi-agent system. As shown in Fig.~\ref{fig:3D_Design}, $\mathfrak{g}_i$ attempts to rotate $\bm{h}_i$ to align with $\bm{f}_i$, and $\mathfrak{u}_i$ represents the magnitude of $\bm{f}_i$ on the projection of $ \bm{h}_i$. The projection-inspired control scheme in Eqs.~\eqref{eq:3D_AffineCtrl1_1}-\eqref{eq:3D_AffineCtrl1_2} is deduced to steer the system to the desired coordinated behaviors.
\end{rema}

Eqs.~\eqref{eq:pos_esti}-\eqref{eq:3D_AffineCtrl1_2} together propose a bearing-based SLAFT control scheme. Since $\bm{p}_i(t)$ is time-varying, the bearing $\bm{g}_{ij} $ and matrix $\bm{\pi}_{\bm{g}_{ij}}$ in Eq.~\eqref{eq:pos_esti} are all time-varying. Before analyzing the stability, we need an assumption such that inter-UAV bearings are well-defined.

\begin{assu}\label{assu:collision}
	No collision occurs between neighboring UAVs during maneuvering, i.e., $\bm{p}_i(t) \ne \bm{p}_j(t) $ for $j \in \mathcal{N}_i$,  $\forall t\ge 0$. 
\end{assu}

Assumption~\ref{assu:collision} imposes a condition widely required in the bearing-only or angle-only collaborative control problems \cite{Chen_2022_SLAF,Cheng_Auto_2024,zhao_bearing-only_2019,Zhang_AST_2022,lin_distributed_2016}. In fact, inter-UAV collisions can be avoided if appropriate initial states are set, and sufficient conditions will be presented later to ensure collision-free system convergence.

For the $i$-th UAV, we define the localization error as $\bm{\delta}_i = \hat{\bm{p}}_i-\bm{p}_i$, and the formation tracking error as $\tilde{\bm{p}}_i = \bm{p}_i - \bm{p}_i^*$.	According to Eq.~\eqref{eq:pos_esti}, the time derivative of the estimation error $\bm{\delta}_i$ can be derived as
\begin{equation}\label{eq:dot_delta}
	\begin{aligned}
		\dot{\bm{\delta}}_i =& \dot{\hat{\bm{p}}}_i-\dot{\bm{p}}_i = -k_\delta \sum_{j \in \mathcal{N}_i^f}\bm{\pi}_{\bm{g}_{ij}}\left(\hat{\bm{p}}_i-\hat{\bm{p}}_j\right) -k_\delta \sum_{j \in \mathcal{N}_i^l}\bm{\pi}_{\bm{g}_{ij}}\left(\hat{\bm{p}}_i-\bm{p}_j\right).
	\end{aligned}	
\end{equation}
Leaders are assumed to move with zero virtual thrust and constant angular speed, that is, $\ddot{\bm{p}}_i = \tau_i\bm{h}_i + v_i \bm{\omega}_i^\wedge \bm{h}_i=v_i \bm{\omega}_r^\wedge \bm{h}_r$ for $i \in \mathcal{V}_l$. Consequently, a follower's desired position $\bm{p}_i^*$ meets the following equation $\ddot{\bm{p}}_i^*=v_i^* \bm{\omega}_r^\wedge \bm{h}_r$. 
Let $\bm{e}_{vi}:= \dot{\tilde{\bm{p}}}_i= \dot{\bm{p}}_i-\dot{\bm{p}}_i^* = \bm{h}_iv_i-\dot{\bm{p}}_i^*$, and the derivative of $\bm{e}_{vi}$ yields
\begin{equation}\label{eq:Dot_Evi1_1}
	\begin{aligned}
		\dot{\bm{e}}_{vi}=&\dot{\bm{h}}_i v_i+ \bm{h}_i \tau_i -\ddot{\bm{p}}_i^*
		=v_i\left(-\dfrac{1}{v_i} \bm{h}_i \times  \bm{f}_i  +\bm{\aleph}_i \right) \times \bm{h}_i  + \bm{h}_i \tau_i -\ddot{\bm{p}}_i^*\\
		=& - \bm{h}_i \times \bm{f}_i \times \bm{h}_i - \bm{h}_i \bm{h}_i^T \bm{f}_i -\ddot{\bm{p}}_i^*  + v_i \bm{\aleph}_i \times \bm{h}_i 
		=\left(\bm{h}_i^\wedge\right)^2  \bm{f}_i- \bm{h}_i \bm{h}_i^T \bm{f}_i = -\bm{f}_i.
	\end{aligned}
\end{equation}

Given $\hat{\bm{p}}_f:=\left[\hat{\bm{p}}_{n_l+1}^T,\cdots,\hat{\bm{p}}_n^T\right]^T \in \mathbb{R}^{dn_f}$, $\bm{\delta}_f:=\left[\bm{\delta}_{n_l+1}^T,\cdots,\bm{\delta}_n^T\right]^T \in \mathbb{R}^{dn_f }$, 
$ \tilde{\bm{p}}_f = \left[ \tilde{\bm{p}}_{n_l+1}^T, \cdots, \tilde{\bm{p}}_n^T\right]^T \in \mathbb{R}^{dn_f }$ and $\bm{e}_{vf} = \left[\bm{e}_{v\left(n_l+1\right)}^T, \cdots, \bm{e}_{vn}^T\right]^T \in \mathbb{R}^{dn_f }$, the compact form of Eqs.\eqref{eq:dot_delta}-\eqref{eq:Dot_Evi1_1} are rewritten as 
\begin{subequations}
	\begin{align}
		\dot{\bm{\delta}}_f =&-k_\delta \bm{L}_{Bfl}\bm{p}_l -k_\delta \bm{L}_{Bff}\hat{\bm{p}}_f = -k_\delta \bm{L}_{Bff}(t)\bm{\delta}_f, \label{eq:Esti_error1}\\
		\dot{\tilde{\bm{p}}}_f =& \bm{e}_{vf}, \\
		\dot{\bm{e}}_{vf} =& - k_p \left(\bar{\bm{\Omega}}_{ff}\hat{\bm{p}}_f+\bar{\bm{\Omega}}_{fl}\bm{p}_l\right)- k_v \left(\bar{\bm{\Omega}}_{ff}\dot{\bm{p}}_f+\bar{\bm{\Omega}}_{fl}\dot{\bm{p}}_l\right) \notag \\
		=& -k_p \left[\bar{\bm{\Omega}}_{ff}\left(\bm{\delta}_f+\tilde{\bm{p}}_f-\bar{\bm{\Omega}}_{ff}^{-1}\bar{\bm{\Omega}}_{fl}\bm{p}_l \right)+\bar{\bm{\Omega}}_{fl}\bm{p}_l\right] - k_v \left(\bar{\bm{\Omega}}_{ff}\dot{\bm{p}}_f+\bar{\bm{\Omega}}_{fl}\dot{\bm{p}}_l\right) \notag\\
		=& -k_p \bar{\bm{\Omega}}_{ff} \tilde{\bm{p}}_f - k_v \bar{\bm{\Omega}}_{ff} \bm{e}_{vf} - k_p \bar{\bm{\Omega}}_{ff}\bm{\delta}_f .
	\end{align}
\end{subequations}

We have the formation tracking error system as below.
\begin{equation}
	\begin{aligned}\label{eq:3D_ErrorSystem}
		\left[\begin{array}{c}
			\dot{\bm{\delta}}_f\\
			\dot{\tilde{\bm{p}}}_f\\
			\dot{\bm{e}}_{vf}
		\end{array}\right] =& \left[\begin{array}{ccc}
			-k_\delta \bm{L}_{Bff}(t) & \bm{0}_{dn_f} & \bm{0}_{dn_f}\\
			\bm{0}_{dn_f} &\bm{0}_{dn_f} & \bm{I}_{dn_f}\\
			-k_p\bar{\bm{\Omega}}_{ff} & -k_p \bar{\bm{\Omega}}_{ff} & -k_v \bar{\bm{\Omega}}_{ff}
		\end{array}\right]\left[\begin{array}{c}
			\bm{\delta}_f\\
			\tilde{\bm{p}}_f\\
			\bm{e}_{vf}
		\end{array}\right].
	\end{aligned}
\end{equation}
The coupled SLAFT dynamics \eqref{eq:3D_ErrorSystem} is complicated because the system matrix is state-dependent and time-varying. Now we present the results on the closed-loop dynamics \eqref{eq:3D_ErrorSystem}.
\begin{theo}\label{theo:3D_AffineTracking}
	Consider a formation with $n$ fixed-wing UAVs described by Eq.~\eqref{eq:3d_unicycle} in three-dimensional space. 
	Under Assumptions~\ref{assump:nominal}-\ref{assu:collision}, the following conclusions hold with the bearing-based SLAFT control scheme Eq.~\eqref{eq:SLAF_ori}:
	\begin{itemize}
		\item[(i)]  The equilibrium set of the dynamics \eqref{eq:3D_ErrorSystem} is 
		\begin{equation}
			\begin{aligned}
				\varXi_{es} =& \left\{ \left(\bm{\delta}_f, \tilde{\bm{p}}_f,   \bm{e}_{vf} \right) \left| \right. \bm{L}_{Bff}(t) \bm{\delta}_f(t) =\bm{0},  \tilde{\bm{p}}_f(t)+ \bm{\delta}_f(t) =\bm{0}, \bm{e}_{vf}(t)=\bm{0} \right\}\\
				=& \left\{ \left(\bm{\delta}_f, \tilde{\bm{p}}_f,   \bm{e}_{vf} \right) \left|  \right. \bm{L}_{Bff}(t) \bm{\delta}_f(t) =\bm{0},  \hat{\bm{p}}_f(t) = \bm{p}_f^*(t) , \bm{e}_{vf}(t)=\bm{0} \right\}.
			\end{aligned}
		\end{equation}

		\item[(ii)] If there exist positive scalars $T$ and $\alpha$ such that for all $\ t >0$
		\begin{equation}\label{eq:PECondition}
			\int_{t}^{t+T}\bm{L}_{Bff}(\tau) d\tau \ge \alpha \bm{I}_{dn_f},
		\end{equation}
		then every solution of the closed-loop dynamics \eqref{eq:3D_ErrorSystem} is bounded, and the origin is exponentially approached by all bounded solutions.
	\end{itemize}
\end{theo}

\begin{proof}
	The proof of Theorem~\ref{theo:3D_AffineTracking} is given in Appendix~\ref{sec:app1}.
\end{proof}

\begin{rema}
	The bearing-based localization algorithm Eq.~\eqref{eq:pos_esti} has a similar form to the localization protocol proposed in \cite{zhao_localizability_2016}, but the main characteristic is that the bearing Laplacian matrix $\bm{L}_{Bff}(t)$ is time-varying and not always nonsingular. The algorithm is not well defined at times when the formation enters a weak observability case. For instance, in a collinear configuration, the system turns into unlocalizable and $\bm{L}_{Bff}(t)$ tends to become singular, resulting in the failure of Eq.~\eqref{eq:SLAF_ori}. To solve the problem, a state-related persistent excitation condition in Eq.~\eqref{eq:PECondition} is needed that assists to encompass the strong and weak observability cases.
	
\end{rema}

The desired equilibrium of system \eqref{eq:3D_ErrorSystem} is the origin $\varXi_{des} =  \left\{ \left(\bm{\delta}_f, \tilde{\bm{p}}_f,   \bm{e}_{vf} \right) \left| \right. \bm{\delta}_f =\bm{0}, \tilde{\bm{p}}_f=\bm{0} , \bm{e}_{vf}=\bm{0}  \right\}$, where the followers converge to their target position with a desired velocity. However, the overall equilibrium set $\varXi_{es}  $ of the dynamics \eqref{eq:3D_ErrorSystem} consists of not only $\varXi_{des}$ but also the undesired equilibrium set $\varXi_{udes} = \varXi_{es} \setminus \varXi_{des} $, where the followers achieve the desired velocity but not reach their target position. Specifically, if the system converges to an unlocalizable configuration, we have $\bm{L}_{Bff}(t)\bm{\delta}_f(t) =\bm{0}$ but $\bm{\delta}_f(t) \ne \bm{0}$ since $\operatorname{rank}(\bm{L}_{Bff}(t)) < dn_f$. Consequently, we propose the persistent excitation condition \eqref{eq:PECondition} in Theorem~\ref{theo:3D_AffineTracking} to ensure the localizability of the coupled system, which lays the foundation for analyzing the convergence of the localization and formation tracking errors. Since $\bm{L}_{Bff}(t)$ is symmetric and positive semi-definite, it is not necessary for $\bm{L}_{Bff}(t)$ to be positive definite throughout the entire duration $T$ to ensure that the condition \eqref{eq:PECondition} is satisfied. That is, if the formation is localizable at certain moments, it is sufficient to guarantee that $\int_{t}^{t+T} \bm{L}_{Bff}(\tau) d\tau - \alpha \bm{I}_{dn_f} > 0$, as verified in the simulation.

Nevertheless, it is hard to maintain the state-related Eq.~\eqref{eq:PECondition} in some special situations, such as when the collinear configuration lasts for a period of time. To complete the control scheme, we propose a perturbation-based approach to avoid falling into unlocalizable configurations while relaxing the dependence on the persistent excitation condition \eqref{eq:PECondition}, inspired by \cite{Trinh_TAC_2020,Chen_2022_SLAF}.

\begin{rema}\label{rema:2D}
	Following a similar approach, the control strategy proposed in Eq.~\eqref{eq:SLAF_ori} can be applied in two-dimensional unicycle model,
	\begin{equation}
		\left\{\begin{array}{l}
			\dot{x}_i = v_i \cos \theta_i,\\
			\dot{y}_i = v_i \sin \theta_i,\\
			\dot{\theta}_i = \omega_i,\\
		\end{array}\right.\notag
	\end{equation}
	where $\theta_i$ represents the heading angle of the $i$-th UAV, $v_i$ and $\omega_i$ are the corresponding control inputs. In particular, we can apply the localization method Eq.~\eqref{eq:pos_esti} while redefining the affine formation control laws based on our previous work  \cite{LHM_FITEE}. Let $\bm{h}_i = \left[\begin{array}{cc}
		\cos \theta_i & \sin \theta_i\\	
	\end{array}\right]^T$ and $\bm{h}_i^\perp = \left[\begin{array}{cc}
		-\sin \theta_i & \cos \theta_i\\	
	\end{array}\right]^T$, and redefine $\bm{f}_i :=  \sum_{j \in \mathcal{N}_i^f}k_p \varpi_{ij}  \left(\hat{\bm{p}}_i-\hat{\bm{p}}_j\right)+ \sum_{j \in \mathcal{N}_i^l}k_p \varpi_{ij}  \left(\hat{\bm{p}}_i-\bm{p}_j\right) + \dot{\bm{p}}_i^* +\omega_r \bm{h}_i^{\perp}$. Then, we have the following dynamics,
	\begin{subequations}\label{eq:2D_LAF}
		\begin{align}
			\dot{\hat{\bm{p}}}_f =&  -k_\delta \bm{L}_{Bff}(t)\bm{\delta}_f + \dot{\bm{p}}_f, \\
			\bm{v}_f =& \bm{D}_h \left(-\bar{\bm{\Omega}}_{ff} \hat{\bm{p}}_f - \bar{\bm{\Omega}}_{fl}  \bm{p}_l + \dot{\bm{p}}_f^*\right),  \\
			\bm{\omega}_f =& \bm{D}_{h^\perp} \left(-\bar{\bm{\Omega}}_{ff} \hat{\bm{p}}_f - \bar{\bm{\Omega}}_{fl}  \bm{p}_l + \dot{\bm{p}}_f^*\right) +\omega_r  \otimes \bm{1}_{n_f} ,
		\end{align}
	\end{subequations}
	where $\bm{D}_h = \operatorname{diag}\left(\bm{h}_{n_l+1}^T,~\cdots,~\bm{h}_n^T\right)$ and $\bm{D}_{h^\perp} = \operatorname{diag}\left(\bm{h}_{n_l+1}^{\perp T},~\cdots,~\bm{h}_n^{\perp T}\right)$. With a persistent excitation condition \eqref{eq:PECondition}, the stability analysis is similar to the proof in Appendix~\ref{sec:app1} and \cite{LHM_FITEE}. Specifically, the limiting equation theorem shown in \cite{Itzhak_Defending_2014} is essential to prove that the convergence of $\tilde{\bm{p}}_f $ is not affected by the transient term $\bm{\delta}_f$ when Eq.~\eqref{eq:PECondition} holds. Consequently, the equilibrium set $\left\{ \left(\bm{\delta}_f, \tilde{\bm{p}}_f \right) \left| \right. \bm{\delta}_f =\bm{0}, \tilde{\bm{p}}_f=\bm{0}  \right\}$ is asymptotically approached. Due to page limitations, detailed theoretical analysis is omitted here.
	
\end{rema}

\subsection{The perturbation-based SLAFT control scheme}

To ensure that the system can escape away from undesired configurations in $\varXi_{udes}$ while relaxing the state-related persistent excitation condition in Eq.~\eqref{eq:PECondition}, we modify the localization and affine formation tracking control laws as follows,
\begin{subequations}\label{eq:LAF_per}
	\begin{align}
		\dot{\hat{\bm{p}}}_i =& -k_\delta \sum_{j \in \mathcal{N}_i^f}\bm{\pi}_{\bm{g}_{ij}}\left(\hat{\bm{p}}_i-\hat{\bm{p}}_j\right) -k_\delta \sum_{j \in \mathcal{N}_i^l}\bm{\pi}_{\bm{g}_{ij}}\left(\hat{\bm{p}}_i-\bm{p}_j\right)  + \sum_{j \in \mathcal{N}_i} k_p \varpi_{ij}  \left(\dot{\bm{p}}_i - \dot{\bm{p}}_j \right) + \dot{\bm{p}}_i  ,  \\
		\tau_i =& \bm{h}_i^T \left[ - \bm{f}_i - \bm{\eth}_i\left(t\right) \right],  \\
		\bm{\omega}_i =&\dfrac{1}{v_i} \bm{h}_i \times  \left[ - \bm{f}_i -\bm{\eth}_i\left(t\right) \right] + \bm{\aleph}_i ,
	\end{align}
\end{subequations}
where $\bm{\eth}_i\left(t\right) = \varsigma_i(t)\left(\operatorname{sgn}\left(\dot{\bm{p}}_i - \dot{\bm{p}}_i^* \right) - \bm{\kappa}_i\left(t\right)  \right)$. The function $\operatorname{sgn}\left(\cdot\right)$ represents the signum function. $\varsigma_i(t) = \sum_{j \in \mathcal{N}_i} \| \hat{\bm{g}}_{ij}\left(t\right) - \bm{g}_{ij}\left(t\right)\|_2 $ is the sum of the estimated bearing errors associated with UAV $i$. $\bm{\kappa}_i\left(t\right)$ is a continuous time-varying vector satisfying $\|\bm{\kappa}_i\left(t\right) \|_2 <1$. 

\begin{theo}\label{theo:3D_AffineTracking_per}
	Consider a multi-agent system consisting of $n$ fixed-wing UAVs governed by Eq.~\eqref{eq:3D_unicycle_matrix}. Under Assumptions~\ref{assump:nominal}\mbox{-}\ref{assu:collision} and the perturbation-based SLAFT control scheme \eqref{eq:LAF_per}, if the desired formation is localizable, the multi-agent system globally asymptotically converges to the desired equilibrium $\varXi_{des}$.
\end{theo}

\begin{proof}
	The proof of Theorem~\ref{theo:3D_AffineTracking_per} is omitted.
\end{proof}

\begin{rema}
	The perturbation term $\bm{\eth}_i\left(t\right)$ in Eq.~\eqref{eq:LAF_per} has a positive effect of driving the system to the desired configurations, but it also increases the convergence time of the coupled system, and the exponential convergence shown in Theorem \ref{theo:3D_AffineTracking} cannot be guaranteed. Moreover, to meet $\| \bm{\kappa}_i \|_2 < 1$ and guarantee that the perturbation term is small enough, we can choose  $ \bm{\kappa}_i(t) = e^{-\gamma_i t} \left[\begin{array}{ccc}
		\frac{\sqrt{2}}{2} \cos(t) & \frac{\sqrt{2}}{2} \cos(t) &  \sin(t) 
	\end{array}
	\right]^T$ where $\gamma_i$ is a positive scalar.
\end{rema}

\begin{rema}
	Unlike the simultaneous localization and formation algorithms developed in \cite{GUO_IFAC_2020,Nguyen_TRO_2020,Guo_TCY_2020,Fang_TAC_2024}, which rely on inter-agent displacement or distance measurements, we investigate the bearing-based SLAFT control problem in this paper, where localization and affine formation control are carried out simultaneously. Both constant and time-varying inter-UAV bearings caused by the affine transformation can be achieved with the proposed integrated control scheme. A key outcome of our work is  proposing the SLAFT control scheme for the nonholonomic systems in the three-dimensional space,  in contrast to the most frequently-used simplified model (i.e., the linear integrator UAV model) in the relevant literature \cite{ye_bearing-only_2017,tang_localization_2023,Chen_2022_SLAF,Fang_TAC_2024}. 
\end{rema}

Next, we introduce a sufficient condition on the initial conditions for the closed-loop system to avoid collisions between the neighbors. 
\begin{prop}\label{prop:AvoidCollision}
	Under Assumptions~\ref{assump:nominal}-\ref{assu:Dp}, for $\forall t \ge 0,~i \ne j,~j\in \mathcal{N}_i$, if the following condition holds,
	\begin{equation}\label{eq:IntialCondition}
		\begin{aligned}
			&\frac{k_p \lambda_{min} \left(\bar{\bm{\Omega}}_{ff}\right)}{2 n_f}\left(\|\bm{e}_{ij}^*(t) \|-\iota\right)^2 \ge   \| \bm{\delta}_f(0)\| ^2 +  k_p \tilde{\bm{p}}_f(0)^T \bar{\bm{\Omega}}_{ff} \tilde{\bm{p}}_f(0) +  \| \bm{e}_{vf}(0)\|^2 ,
		\end{aligned}
	\end{equation}
	the solution of the closed-loop system satisfies
	\begin{equation} \label{eq:CollisionAvoidance}
		\|\bm{e}_{ij}(t) \| \ge \iota,
	\end{equation}
	where $\iota$ is the safety distance between any two neighboring UAVs satisfying $ \inf_{t \ge 0,~i\ne j} \|\bm{e}_{ij}^* (t)\| \ge \iota > 0$.
\end{prop}

\begin{proof}
	Based on the initial condition \eqref{eq:IntialCondition} and Lyapunov function \eqref{eq:Lyapunov}, we have
	\begin{equation}
		V(0) \le \frac{k_p \lambda_{min} \left(\bar{\bm{\Omega}}_{ff}\right)}{4 n_f}\left(\|\bm{e}_{ij}^*(t) \|-\iota\right)^2,~\forall t \ge 0.
	\end{equation}
	Combined with Theorem~\ref{theo:3D_AffineTracking_per}, it is noted that 
	
	\begin{equation}
		\begin{aligned}
			\| \tilde{\bm{p}}_f(t) \|^2 \le & \frac{1}{\lambda_{min} \left(\bar{\bm{\Omega}}_{ff}\right)} \tilde{\bm{p}}_f^T(t) \bar{\bm{\Omega}}_{ff} \tilde{\bm{p}}_f(t) 
			\le  \frac{1}{ k_p \lambda_{min} \left(\bar{\bm{\Omega}}_{ff}\right)} V(t) \le \frac{1}{ k_p \lambda_{min} \left(\bar{\bm{\Omega}}_{ff}\right)} V(0)
			\le  \frac{1}{4 n_f}\left(\|\bm{e}_{ij}^*(t) \|-\iota\right)^2.
		\end{aligned}
	\end{equation}
	
	Accordingly, it is deduced that 
	\begin{equation}
		\begin{aligned}
			\|\bm{e}_{ij}(t) \|  =& \|\bm{p}_j(t) - \bm{p}_i(t) \| =  \|  \tilde{\bm{p}}_j(t) - \tilde{\bm{p}}_i(t) + \bm{e}_{ij}^*(t) \| 
			\ge  \|\bm{e}_{ij}^*(t)\| - \|\tilde{\bm{p}}_i(t)\| -\|\tilde{\bm{p}}_j(t)\| 
			\ge  \|\bm{e}_{ij}^*\| - 2\sqrt{n_f} \| \tilde{\bm{p}}_f\|
			\ge  \iota,
		\end{aligned}
	\end{equation}
	which together with \eqref{eq:IntialCondition} implies \eqref{eq:CollisionAvoidance}.
\end{proof}
Proposition~\ref{prop:AvoidCollision} indicates that Assumption \ref{assu:collision} can be guaranteed through initial conditions, which is achievable in practice.

\section{SIMULATION} \label{sec:Simulation}
In this section, simulations are carried out to validate the effectiveness of the proposed control strategies in two\mbox{-} and three\mbox{-}dimensional spaces. 

\begin{figure}
	\centering
	\includegraphics[scale=0.25]{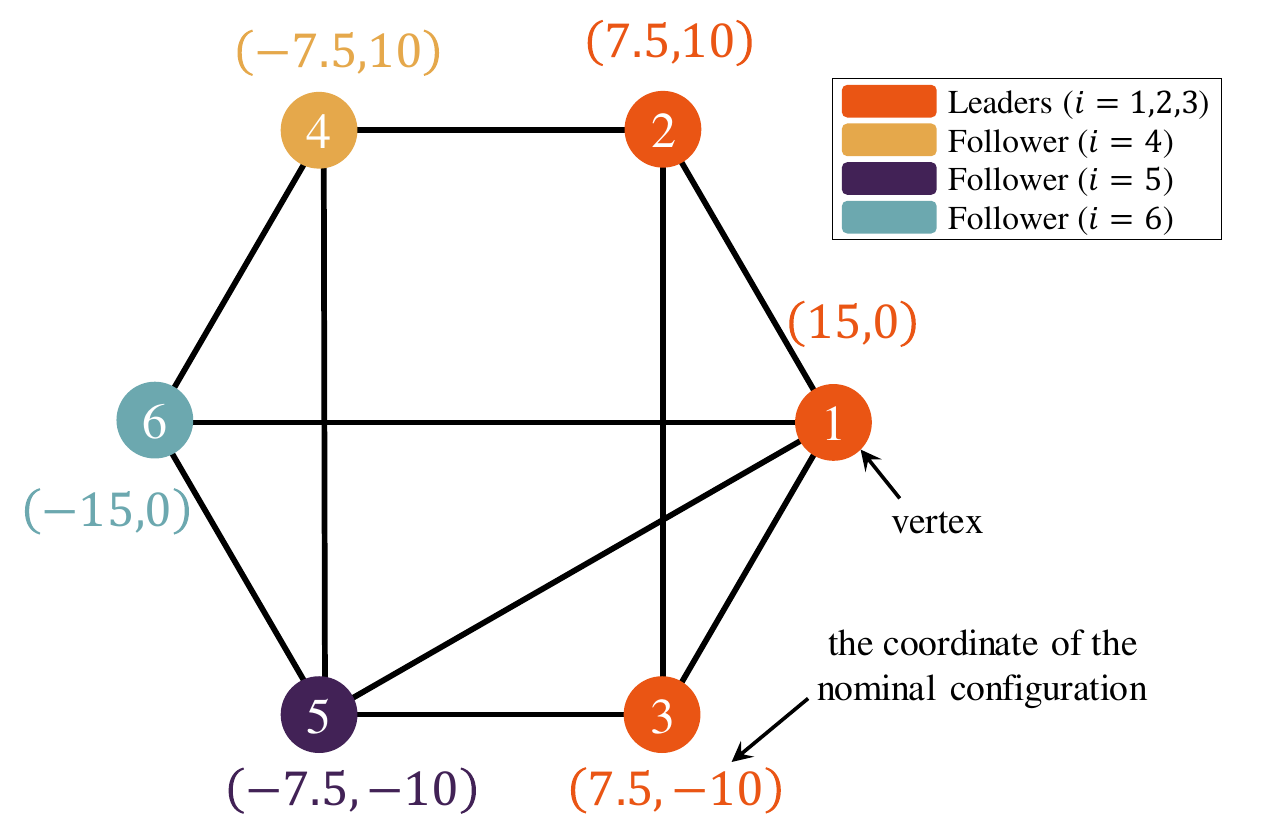} 
	\caption{The interaction topology of fixed-wing UAVs in two-dimensional space.}
	\label{fig:2D_Topology}
\end{figure}

\begin{figure*}
	\centering
	\includegraphics[scale=0.55]{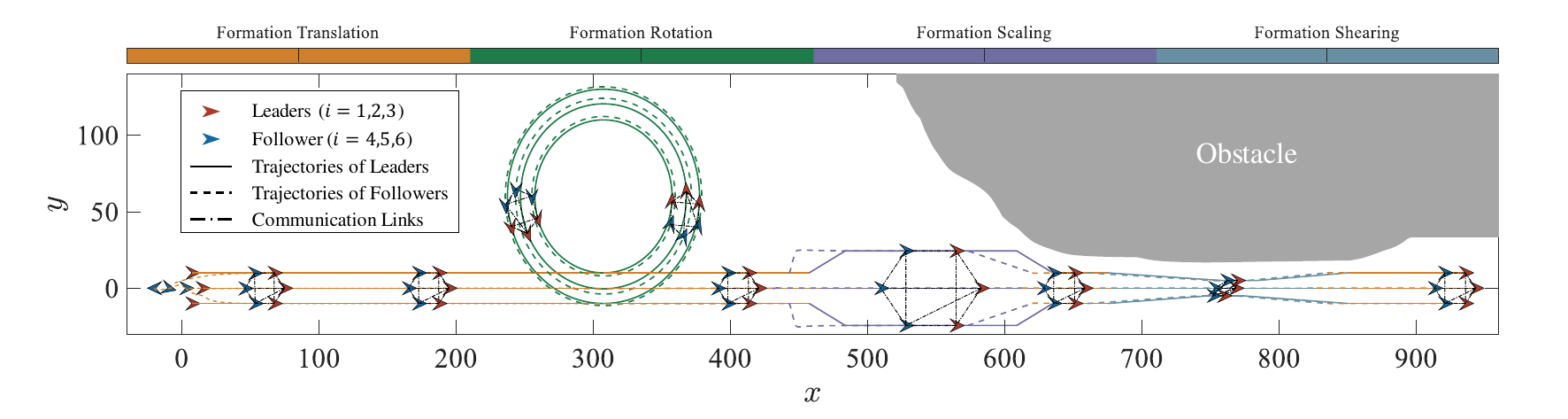} 
	\caption{Trajectories of six UAVs achieving target affine formations in two-dimensional space. Different colors in the trajectories represent the different stages of the affine transformation. The red wedge-shape icons indicate the leaders and the remaining three dark cyan wedge-shape icons are the followers. The dotted dashed lines among the UAVs represent the information connections.}
	\label{fig:2D_Traje}
\end{figure*}

\begin{figure}
	\centering
	\includegraphics[scale=0.4]{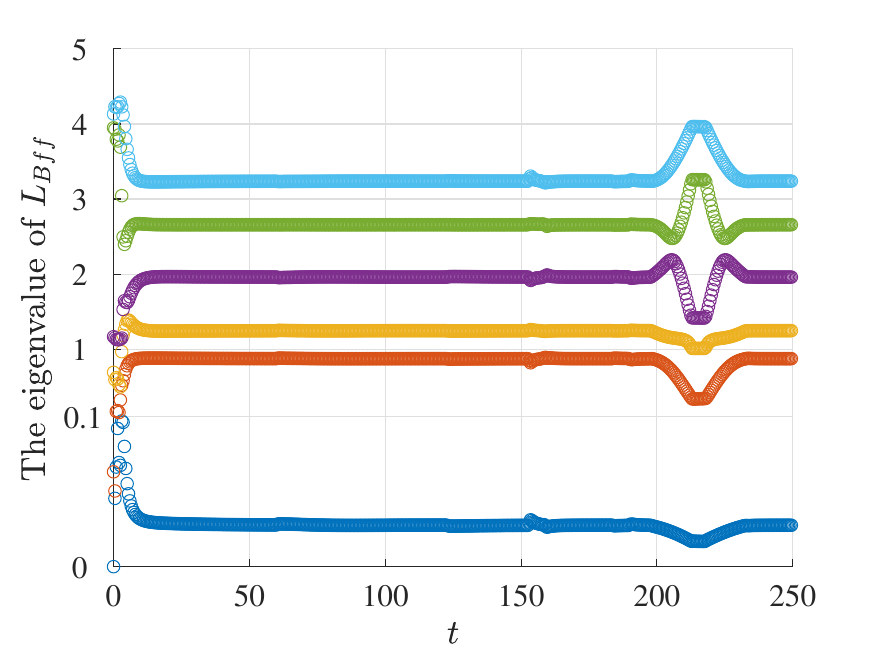} 
	\caption{Time evolution of the six eigenvalues of $\bm{L}_{Bff}(t)$, which are arranged in ascending order and distinguished by color.}
	\label{fig:2D_Eigenvalue}
\end{figure}

\begin{figure*}
	\centering
	\subfigure[Time evolution of the norm of the bearing-based localization error $\| \delta_i \|$]{\includegraphics[scale=0.32]{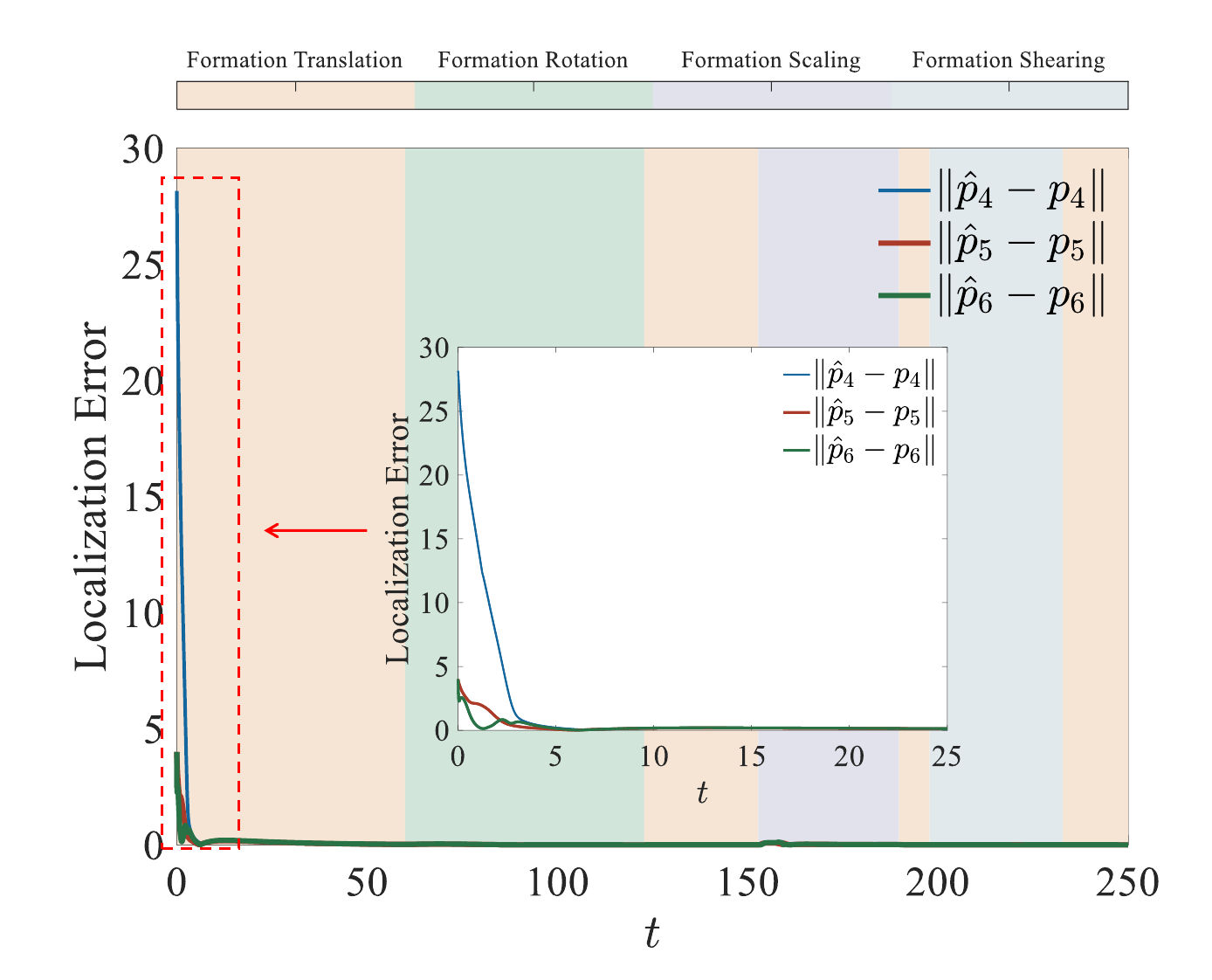} }
	\subfigure[Time evolution of the norm of the affine formation tracking error $\| \tilde{p}_i \|$]{\includegraphics[scale=0.32]{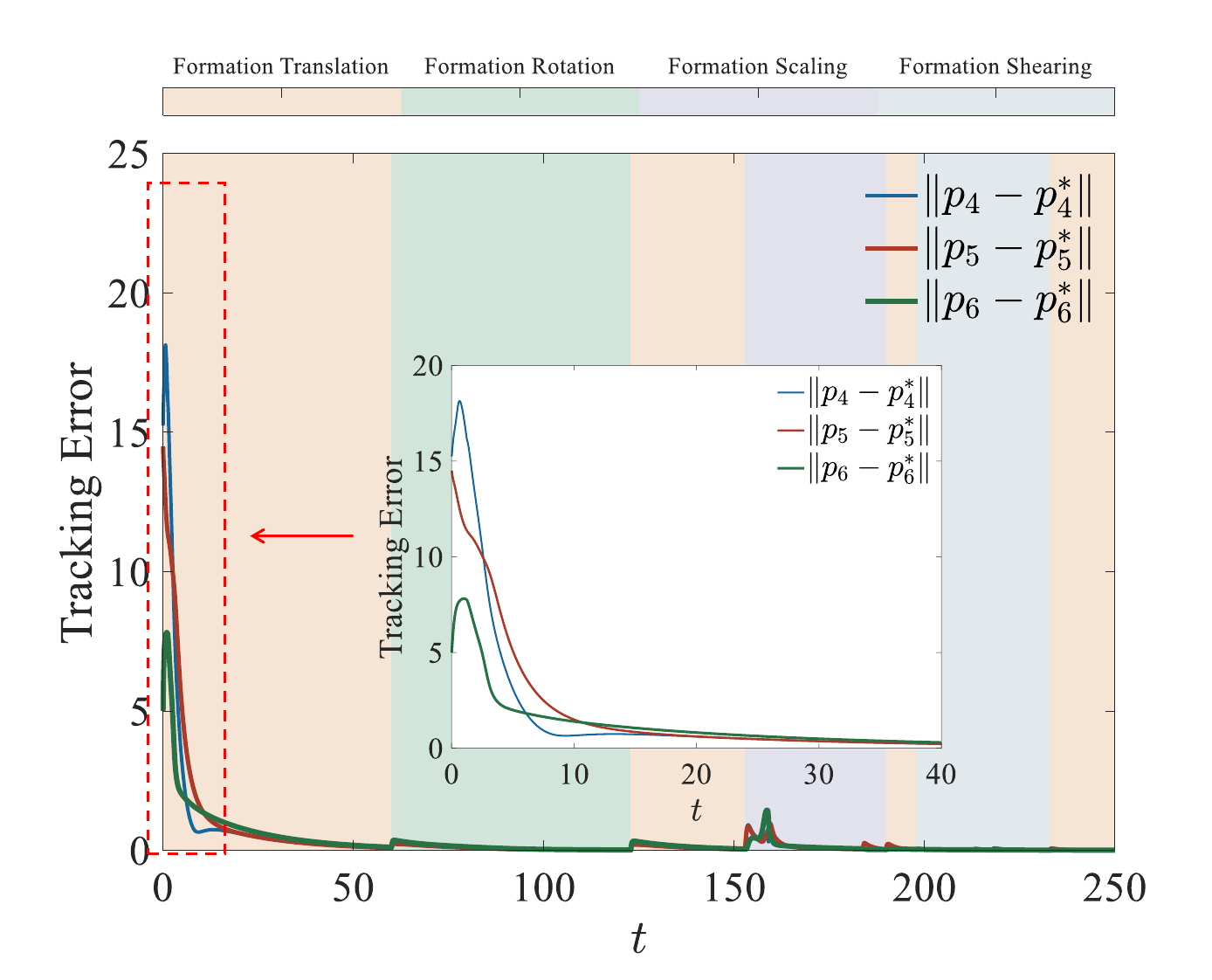} }
	\caption{Time evolution of the bearing-based localization error and affine formation tracking error for followers. The shaded areas with colors represent different stages of the affine transformation.}
	\label{fig:2D_TrackLolization_Error}
\end{figure*}

We first consider a scenario in two-dimensional space with system dynamics described by Eq.~\eqref{eq:2D_LAF}. The underlying graph is designed as shown in Fig.~\ref{fig:2D_Topology}, where the first three UAVs are chosen as leaders. The eigenvalues of the corresponding stress matrix satisfy $\lambda\left(\bm{\Omega}\right) = \left\{0,0,0,0.788,2.3641,2.3641\right\}$ and $\lambda\left(\bm{\Omega}_{ff}\right) = \left\{0.0676,0.3940,2.2965\right\}$, implying that the nominal formation is affine localizable and universally rigid. Fixed-wing UAVs are designed to translate along a straight line as shown in Fig.~\ref{fig:2D_Traje}, where the colored line segments represent different stages of affine transformations including formation rotation, scaling, shearing and translation. Fig.~\ref{fig:2D_Eigenvalue} presents the evolution of the eigenvalue of the bearing Laplacian matrix $\bm{L}_{Bff}(t)$. It is worth noting that in the initial state, the three followers are collinear such that $\bm{L}_{Bff}(0)$ is not positive definite. As analyzed earlier, the PE condition \eqref{eq:PECondition} does not require the formation to be localizable all the time, and the PE condition \eqref{eq:PECondition} is satisfied in our simulation. As can be observed from Fig.~\ref{fig:2D_TrackLolization_Error}, the bearing-based localization and formation tracking errors converge to zero, which means $\bm{p}_{i}(t) \to \bm{p}_{i}^*(t)$ so that the inter-UAV bearings $\bm{g}_{ij}(t)$ converge to $\bm{g}_{ij}^*(t)$ under the proposed SLAFT control scheme. In particular, the inter-UAV bearings $\bm{g}_{ij}^*$ are not always constant during affine transformations, such as formation rotations, demonstrating the effectiveness of our method in tracking time-varying bearings.  

\begin{figure}
	\centering
	\includegraphics[scale=0.25]{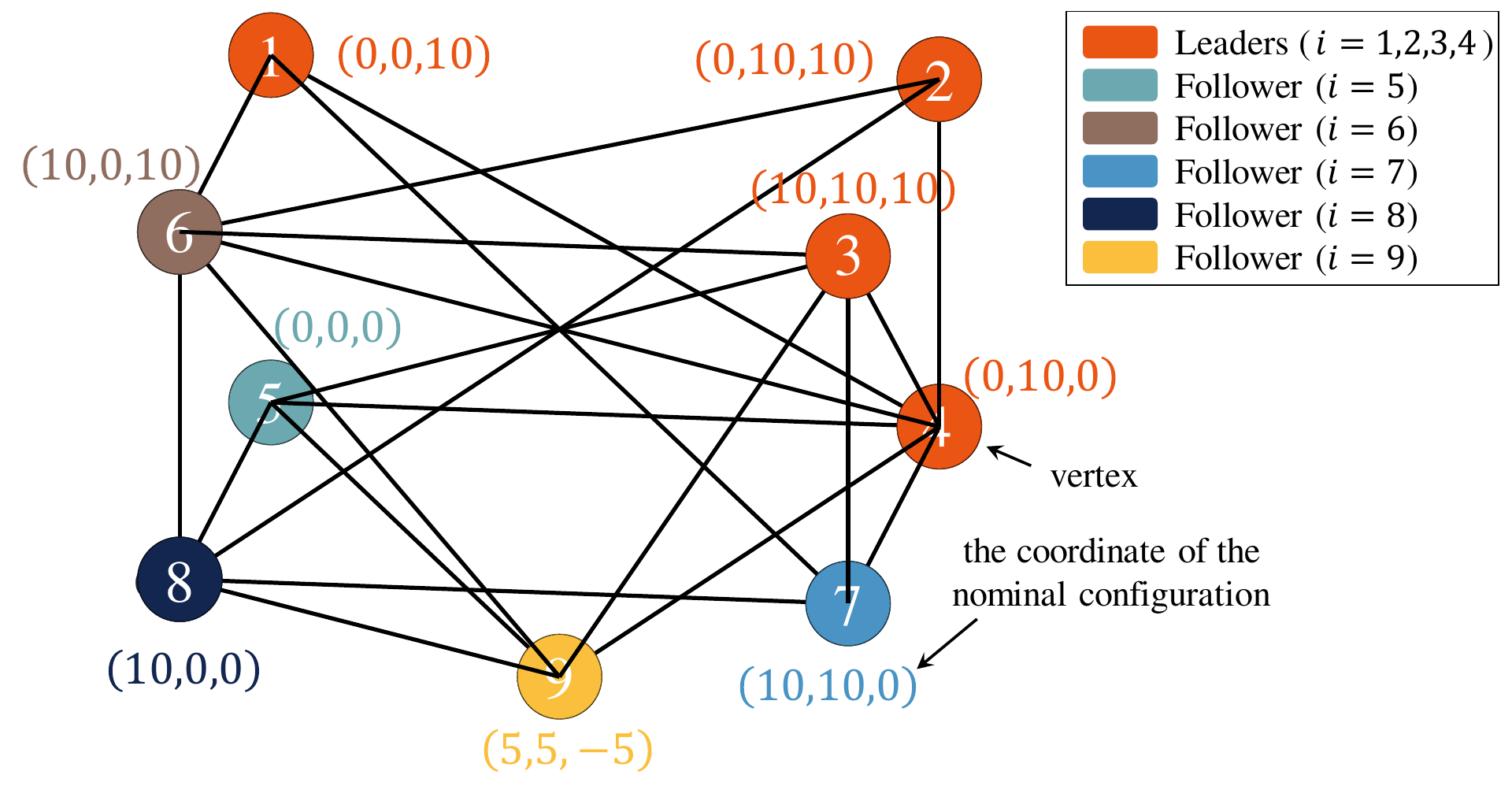} 
	\caption{The interaction topology of fixed-wing UAVs in three-dimensional space.}
	\label{fig:3D_Topology}
\end{figure}

\begin{figure}
	\centering
	\subfigure[Trajectories in 3D space, from the front view]{\includegraphics[scale=0.28]{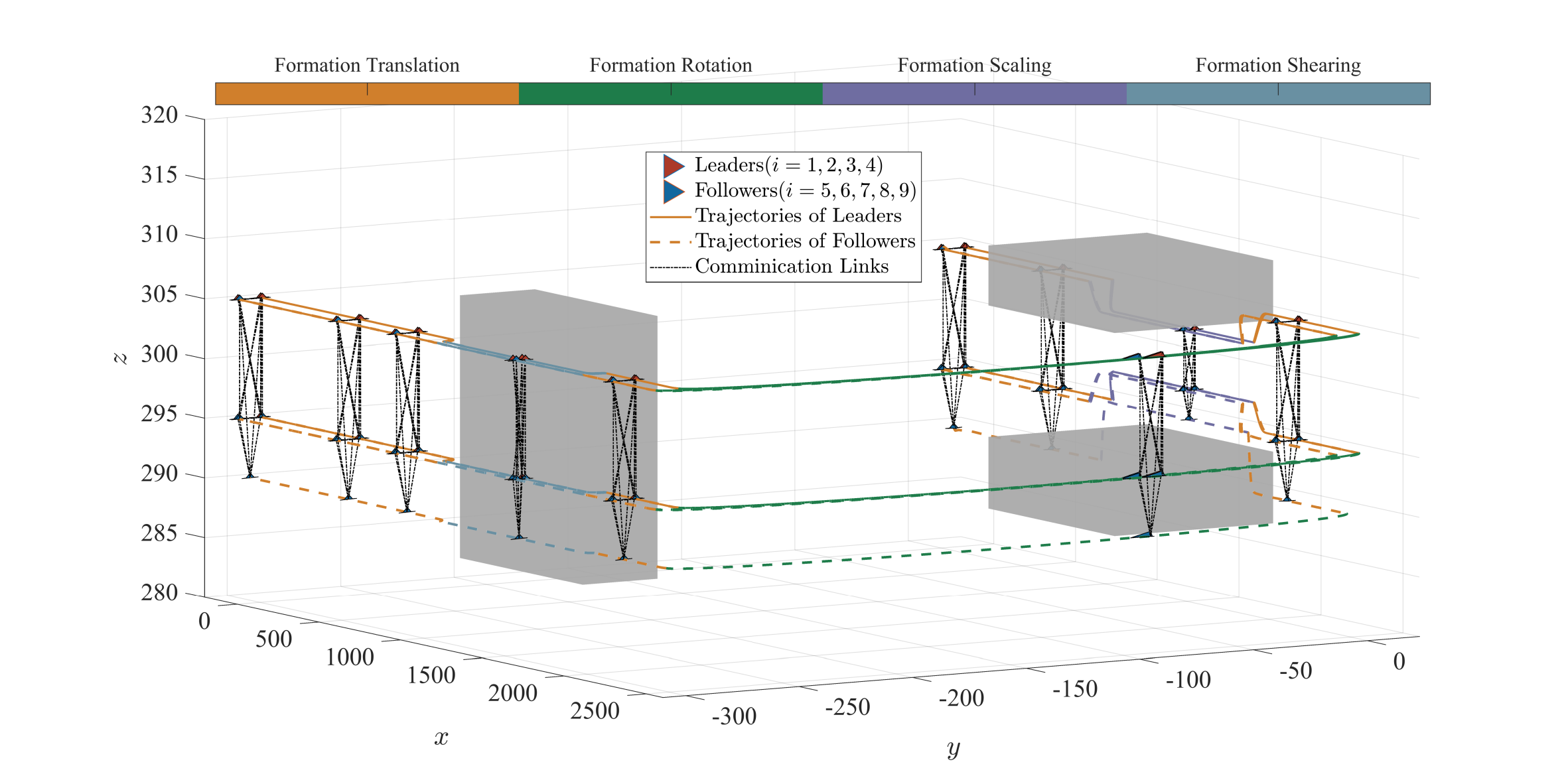} 
	}
	\subfigure[Trajectories in 3D space, from the back view]{\includegraphics[scale=0.2]{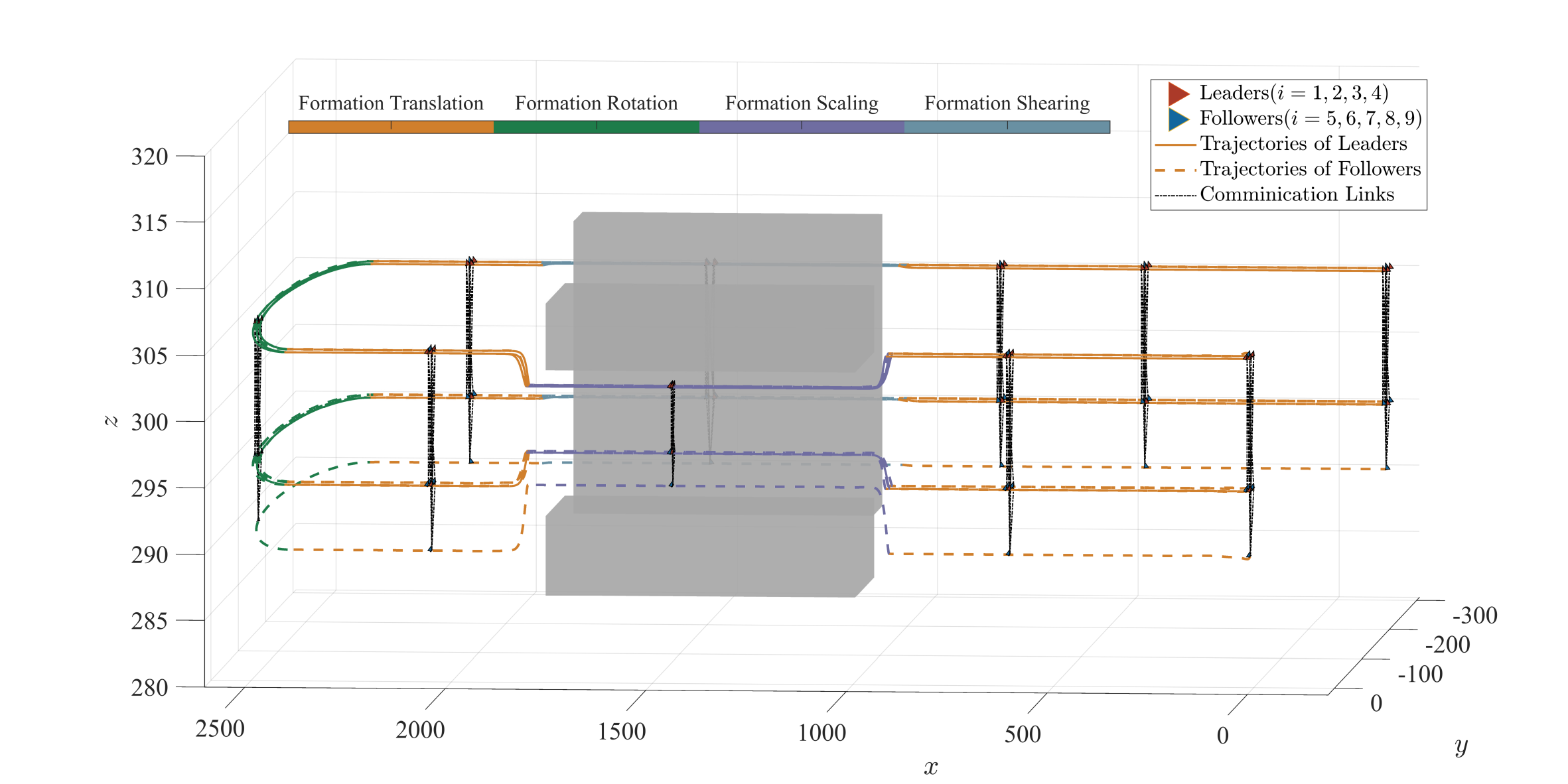} 
	}
	\subfigure[Trajectories in $x-y$ plane]{\includegraphics[scale=0.2]{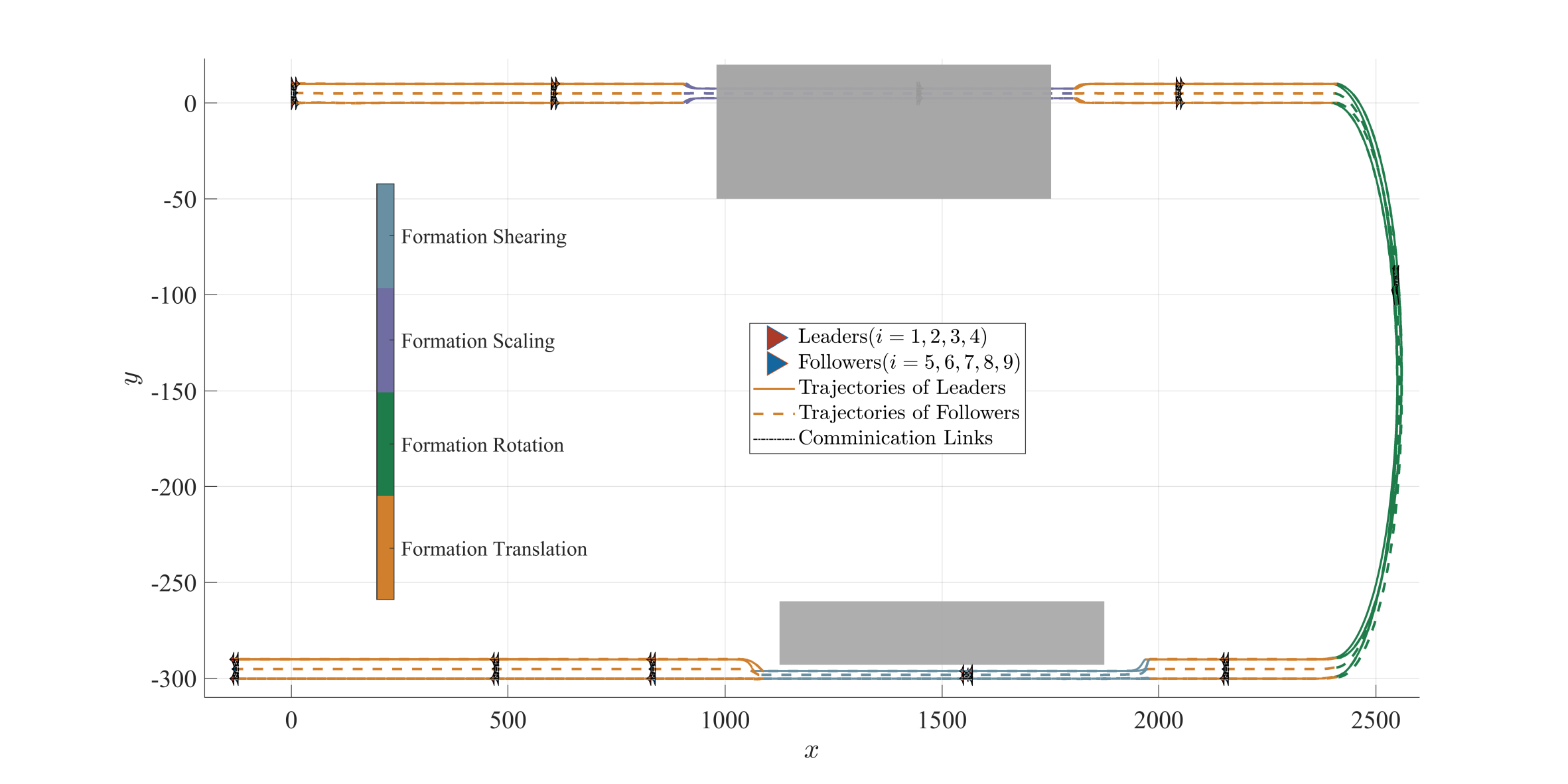} 
	}
	\caption{Trajectories of nine UAVs achieving target affine formations in three-dimensional space. Different colors in the trajectories represent the different stages of the affine transformation. The red wedge-shape icons indicate the leaders and the remaining five dark cyan wedge-shape icons are the followers. The dotted dashed lines among the UAVs represent the topology connections.}
	\label{fig:3D_Trajectory}
\end{figure}

\begin{figure}
	\centering
	\subfigure[Time evolution of the norm of the bearing-based localization error $\| \delta_i \|$]{\includegraphics[scale=0.32]{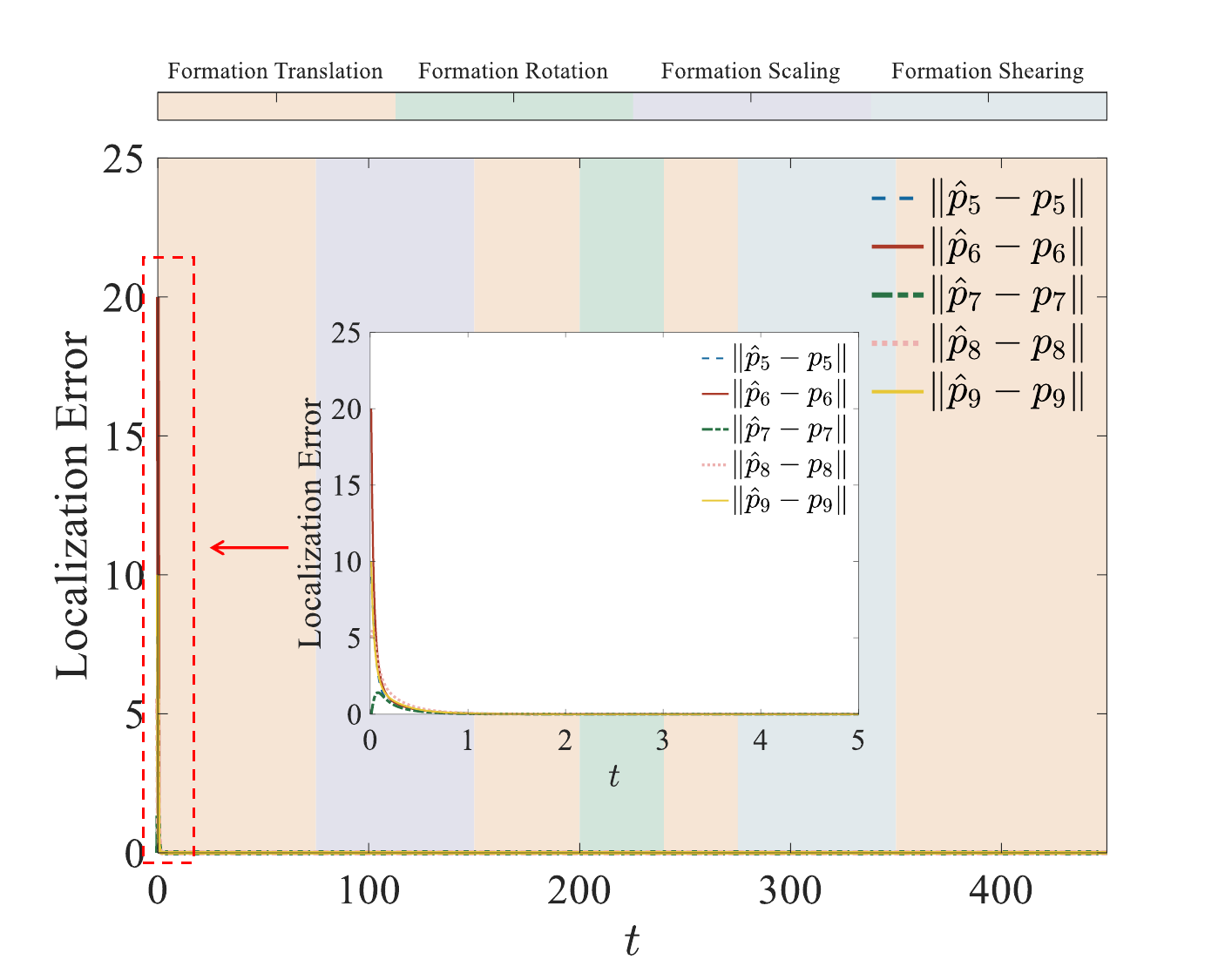} }
	\subfigure[Time evolution of the norm of the affine formation tracking error $\| \tilde{p}_i \|$]{\includegraphics[scale=0.32]{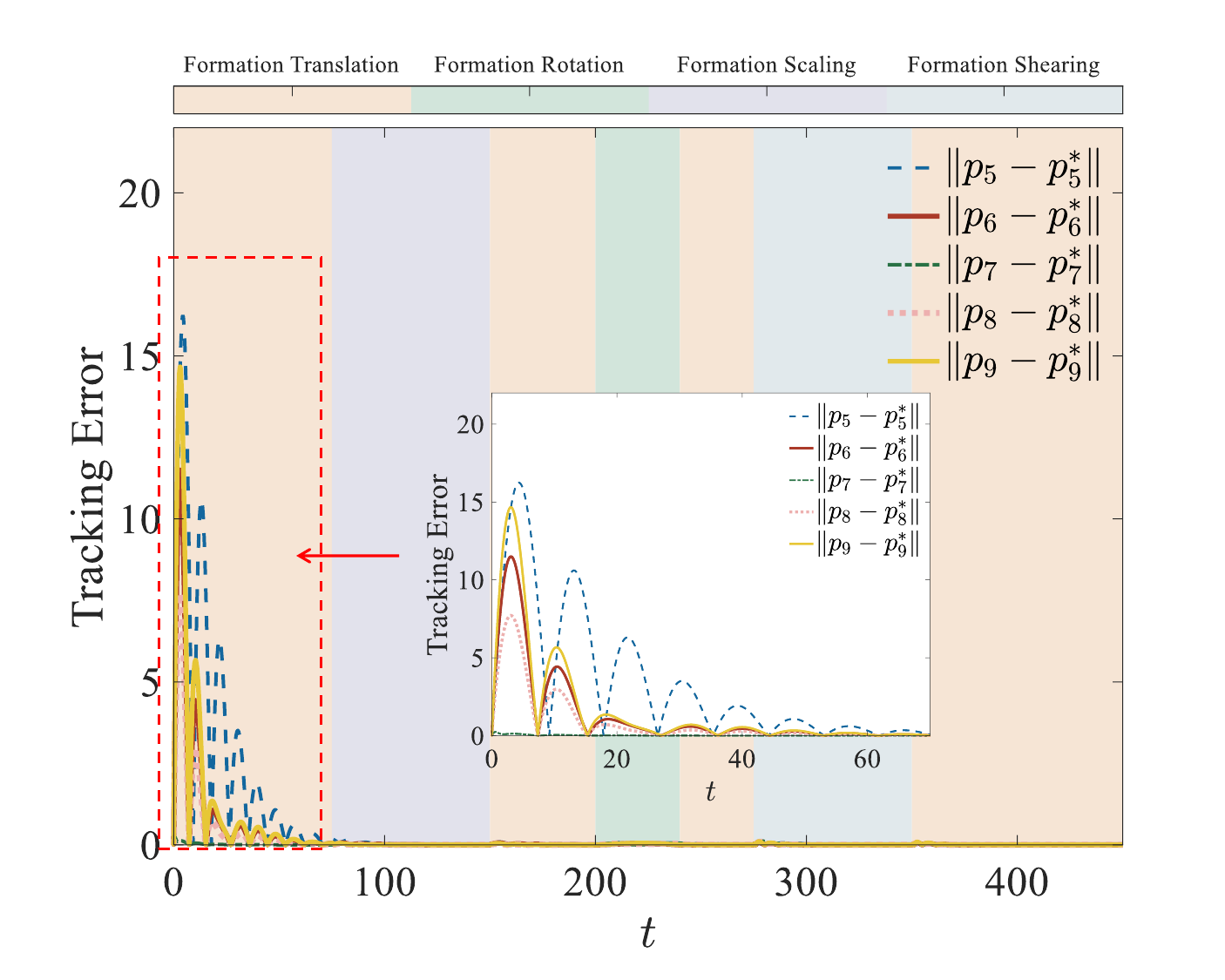} }
	\caption{Time evolution of the bearing-based localization error and affine formation tracking error for followers. The shaded areas with colors represent different stages of the affine transformation.}
	\label{fig:3D_TrackLocalization_Error}
\end{figure}

The second simulation is implemented in a three\mbox{-}dimensional scenario with the perturbation-based control protocol \eqref{eq:LAF_per}. Four UAVs are set as leaders, and the underlying graph is designed as in Fig.~\ref{fig:3D_Topology}. 
The eigenvalues of the corresponding stress matrix satisfy $\lambda\left(\bm{\Omega}\right) = \left\{0,0,0,0,0.0129,0.225,0.486,1.195,1.622\right\}$ and $\lambda\left(\bm{\Omega}_{ff}\right) = \left\{0.0087,0.016,0.3,0.468,1.32\right\}$. Similarly, to illustrate the proposed affine formation control scheme in  three-dimensional space, the target maneuvering including translation, shearing, scaling and rotation is pre-planned by the motion planner.
The trajectories of UAVs are depicted in Fig.~\ref{fig:3D_Trajectory}. It can be seen that the desired affine formation is achieved under the proposed control scheme. The localization and formation tracking errors converge to zero as proved in Theorem~\ref{theo:3D_AffineTracking_per}, as depicted in Fig.~\ref{fig:3D_TrackLocalization_Error}.  
These results demonstrate the SLAFT method can achieve affine transformations in three-dimensional space, even with time-varying desired bearings. 

\section{CONCLUSIONS}\label{sec:Conclusion}
This paper has presented novel bearing-based SLAFT control schemes to handle the time-varying affine formation tracking problem for fixed-wing UAVs. Especially, a perturbation-based control scheme is proposed to prevent UAVs from falling into undesired equilibrium points. With our control strategies, distributed bearing-based self-localization methods are designed and observer-based controllers are proposed to track target affine formations with constant and time-varying inter-UAV bearings. We have provided comprehensive technical analysis and simulations to demonstrate the effectiveness of coupled systems. One future research direction is to adapt the proposed SLAFT scheme to switching graph topologies, since time-varying interaction is ubiquitous for multiple UAV networks. Furthermore, the theoretical analysis and validation of a cooperative localization and formation tracking system based on realistic bearing sensors is an interesting research problem, including observability, stability and flight experiments. 

\clearpage

\appendix
\section{The proof of Theorem~\ref{theo:3D_AffineTracking}}\label{sec:app1}
The following lemma is introduced for the convergence analysis of the overall multi-agent system.

\begin{lemm}[$\lambda-$UGES \cite{Loria_CDC_2000}]\label{lemma:UGES}
	Let $W\left(\cdot, \cdot \right) : \mathbb{R}^+ \times \mathcal{D} \to \mathbb{R}^{n \times r} $ be a continuous function, where $ \mathcal{D} \in \mathbb{R}^q$ is a closed not necessarily compact set. For the system 
	\begin{equation}\label{eq:PE}
		\dot{x}(t) = -W(t,\lambda)W(t,\lambda)^T x(t),
	\end{equation} 
	where $\lambda \in \mathcal{D}$ is a parameter for the system, assume that there exists a constant $W_M >0$ such that for all $t \ge 0$ and $\lambda \in \mathcal{D}$,	$\|W(t,\lambda) \| \le W_M$. If there are two positive parameters $\alpha>0$ and $T > 0$ such that 
	\begin{equation}
		\int_{t}^{t+T}W(\tau,\lambda)W(\tau,\lambda)^T d\tau \ge \alpha \bm{I}_{n},~~\forall t \ge 0
	\end{equation} 
	the system \eqref{eq:PE} is $\lambda-$uniformly globally exponentially stable ($\lambda-$UGES). That is, there exists a constant $\gamma_\lambda \ge \alpha / T\left(1+W_M^2 T\right)^2$ such that $\| x(t,\lambda,t_0,x_0) \| \le \|x_0 \| e^{-\gamma_\lambda \left(t-t_0\right)} $ for all $t \ge t_0$.
\end{lemm}

Different from the widely used theorem proposed in \cite{Anderson_PE_1986}, Lemma~\ref{lemma:UGES} states the property of persistency of excitation for parameterized systems, and the exponential convergence of systems with parameter $\lambda$ is analyzed. Furthermore, Lemma~\ref{lemma:UGES} can also be used to analyze a nonlinear system along trajectories because the proofs can be carried out verbatim by ignoring the parameter $\lambda$, where $\lambda-$UGES is equivalent to UGES. 
The proof of Theorem~\ref{theo:3D_AffineTracking} is exhibited as below.

\emph{\textbf{Proof of Theorem~\ref{theo:3D_AffineTracking}}}

\emph{(i)} For the linear time-varying system \eqref{eq:3D_ErrorSystem}, let $\dot{\bm{\delta}}_f=\bm{0}$, $\dot{\tilde{\bm{p}}}_f=\bm{0} $ and $ \dot{\bm{e}}_{vf}=\bm{0}$ in Eq.~\eqref{eq:3D_ErrorSystem}, thus one gets
\begin{equation}
\left\{\begin{array}{l}
	-k_\delta \bm{L}_{Bff}(t)\bm{\delta}_f =\bm{0},\\
	\bm{e}_{vf}=\bm{0},\\
	-k_p \bar{\bm{\Omega}}_{ff} \tilde{\bm{p}}_f - k_v \bar{\bm{\Omega}}_{ff} \bm{e}_{vf} - k_p \bar{\bm{\Omega}}_{ff}\bm{\delta}_f=\bm{0},
\end{array}
\right. 
\end{equation}
which implies that $k_p \tilde{\bm{p}}_f + k_p \bm{\delta}_f=\bm{0}$ because $\bm{e}_{vf}=\bm{0}$ and $\bar{\bm{\Omega}}_{ff}$ is positive definite. By using the definition of $\tilde{\bm{p}}_f $ and $\bm{\delta}_f$, the equilibrium set of the system \eqref{eq:3D_ErrorSystem} can be derived as $\varXi_{es}$.

\emph{(ii)}
It is clear that $\bm{\delta}_f$ is independent of $\tilde{\bm{p}}_f $ and $\bm{e}_{vf}$ in the formation tracking error system \eqref{eq:3D_ErrorSystem}. 
To prove that the states (i.e., $\bm{\delta}_f$, $\tilde{\bm{p}}_f $ and $\bm{e}_{vf}$) in Eq.~\eqref{eq:3D_ErrorSystem} converge to the origin exponentially fast, the first step is to show the localization error $\bm{\delta}_f$ converges to zero exponentially. Then, we will show that $\tilde{\bm{p}}_f $ and $\bm{e}_{vf}$ are bounded and go to zero as $t \to \infty$.

It is clarified that $\bm{L}_B(t)$ is symmetric positive semi-definite so that its principal minor, $\bm{L}_{Bff}(t)$, is positive semi-definite. 
Under the condition Eq.~\eqref{eq:PECondition},  
a direct application of Lemma~\ref{lemma:UGES} proves that the system~\eqref{eq:Esti_error1} is UGES, which means $\|\bm{\delta}_f(t)\| \le \| \bm{\delta}_f(t_0) \| e^{-\gamma_\lambda \left(t-t_0\right)}$ with $\gamma_\lambda \ge\frac{\alpha k_\delta}{T \left(1+T L_{BM} \right)^2}$, where $L_{BM}$ is a positive constant satisfying $\|\bm{L}_{Bff}(t) \| \le L_{BM}$, and $\lim\limits_{t \to \infty} \bm{\delta}_f(t) =\bm{0}$ exponentially fast for $0 \le t_0 < t < \infty$.

The formation tracking error dynamics~\eqref{eq:3D_ErrorSystem} can be rearranged as 
\begin{equation}\label{eq:LTI_Error}
\dot{\bm{X}}(t) = \bm{\varPhi}\bm{X}(t)+ \bm{\varDelta}\bm{\delta}_f(t),
\end{equation}
where $\bm{X}(t) := \left[\begin{array}{c}
\tilde{\bm{p}}_f\\
\bm{e}_{vf}
\end{array}\right]$, $\bm{\varPhi} :=\left[\begin{array}{cc}
\bm{0}_{dn_f} & \bm{I}_{dn_f}\\
-k_p \bar{\bm{\Omega}}_{ff} & -k_v \bar{\bm{\Omega}}_{ff}
\end{array}\right] $ and $\bm{\varDelta}: = \left[\begin{array}{c}
\bm{0}_{dn_f} \\
-k_p\bar{\bm{\Omega}}_{ff} 
\end{array}\right]$. Obviously, the state matrix $\bm{\varPhi}$ is Hurwitz and thus the linear time-invariant (LTI) system $\dot{\bm{X}}(t) = \bm{\varPhi}\bm{X}(t)$ is exponentially stable. The solution to the LTI system $\dot{\bm{X}}(t) = \bm{\varPhi}\bm{X}(t)$ is expressed as $\bm{X}(t) = e^{\bm{\varPhi}(t-t_0)}\bm{X}(t_0)$ so that one gets
\begin{equation}\label{eq:Ineq1}
\| \bm{X}(t) \| = \|e^{\bm{\varPhi}(t-t_0)}\bm{X}(t_0) \| \le M \| \bm{X}(t_0) \| e^{-r(t-t_0)},
\end{equation} 
while there exist positive constants $M$ and $r$ such that \eqref{eq:Ineq1} holds \cite{Hsu_Ordinary,Perko_Differential}. Since $k_p$ and $\bar{\bm{\Omega}}_{ff}$ are both bounded, there exists a constant $b>0$ so that $\|  \bm{\varDelta}\bm{\delta}_f(t)  \| \le \|  \bm{\varDelta} \| \| \bm{\delta}_f(t) \| \le b \| \bm{\delta}_f(t_0) \| e^{-\gamma_\lambda \left(t-t_0\right)}$, which means $\bm{\varDelta}\bm{\delta}_f(t)$ is exponentially decaying. Using the similar argument as shown in \cite{Angelo_LTV_1970} (see Theorem 8.3), we will prove that $\bm{X}(t)$ in Eq.~\eqref{eq:LTI_Error} goes to zero  exponentially fast. The solution to Eq.~\eqref{eq:LTI_Error} is 
\begin{equation}
\bm{X}(t) = e^{\bm{\varPhi}(t-t_0)}\bm{X}(t_0) + \int_{t_0}^{t} e^{\bm{\varPhi}(t-\tau)} \bm{\varDelta}\bm{\delta}_f(\tau)  d\tau \notag .
\end{equation}
Therefore, in accordance with \eqref{eq:Ineq1}, we have
\begin{equation}
\begin{aligned}
	\|\bm{X}(t) \| \le& \|e^{\bm{\varPhi}(t-t_0)}\bm{X}(t_0) \| + \|\int_{t_0}^{t} e^{\bm{\varPhi}(t-\tau)} \bm{\varDelta}\bm{\delta}_f(\tau)  d\tau\|\\
	\le& M\| \bm{X}(t_0) \| e^{-r(t-t_0)} + M b \| \bm{\delta}_f(t_0) \| \int_{t_0}^{t} e^{-r(t-\tau)}   e^{-\gamma_\lambda \left(\tau-t_0\right)} d\tau\\
	=&M \| \bm{X}(t_0) \| e^{-r(t-t_0)} +\dfrac{b M\| \bm{\delta}_f(t_0) \|}{r-\gamma_\lambda}\left(e^{-\gamma_\lambda \left(t-t_0\right) }-e^{-r(t-t_0)}\right)\\
	=&\left(M \| \bm{X}(t_0) \| - \dfrac{b M\| \bm{\delta}_f(t_0) \|}{r-\gamma_\lambda}\right) e^{-r(t-t_0)} +\dfrac{b M\| \bm{\delta}_f(t_0) \|}{r-\gamma_\lambda} e^{-\gamma_\lambda \left(t-t_0\right) }.
\end{aligned}
\end{equation}
When $t \to \infty$, we have $e^{-r(t-t_0)} \to 0$ and $e^{-\gamma_\lambda \left(t-t_0\right) } \to 0$ for positive $r$ and $\gamma_\lambda$. Therefore, one concludes that $ \left( \tilde{\bm{p}}_f,   \bm{e}_{vf} \right)=\left(\bm{0},\bm{0} \right) $ is exponentially approached by the solutions of system \eqref{eq:LTI_Error}, i.e., $\tilde{\bm{p}}_f$ and $\bm{e}_{vf}$ converge to zero exponentially. 

To sum up, under the PE condition in \eqref{eq:PECondition}, all solutions of the closed-loop dynamics \eqref{eq:3D_ErrorSystem} are bounded, and the origin is exponentially stable. The proof is completed.


\bibliographystyle{cas-model2-names}

\bibliography{MyRef}

\end{document}